\documentclass[twocolumn]{openjournal}
\newcommand{\tablepage}{}           
\shorttitle{Discrete Chi-square Method (DCM)}
\shortauthors{Jetsu}
\usepackage{color}
\usepackage[hidelinks]{hyperref}
\usepackage[switch,columnwise]{lineno}
\usepackage{skull}    
\usepackage{listings} 
\usepackage{enumitem}
\newcommand{\hilight}[1]{\setlength{\fboxsep}{1pt}\colorbox{yellow}{#1}}
\newcommand{\hlitem}{\stepcounter{enumi}\item[\hilight{\large \theenumi}]}

\newcommand{\kista}[1]{{\color{black} #1}}
\newcommand{\target}{XZ~And}

\newcommand{\PR}[1]{{\color{magenta} {\ttfamily #1}}}

\newcommand{\Webcolour}{\color{blue}}
\newcommand{\Link}[2]{\href{#1}{\Webcolour \bf #2}}

\newcommand{\MyPro}{\PR{dcm.py}}
\newcommand{\MyDat}{\PR{dcm.dat}} 
\newcommand{\MyTest}{\PR{TestData.dat}}

\begin{document}
\title{Discrete Chi-square Method 
for Detecting Many Signals}
\author{L. Jetsu}
\affil{Department of Physics,
  P.O. Box 64,
  FI-00014 University of Helsinki,
  Finland}
\email{lauri.jetsu@helsinki.fi}
\begin{abstract}
  Unambiguous detection of
  signals superimposed on unknown
  trends 
  is difficult
  for unevenly
  spaced data.
  Here, we formulate the Discrete
  Chi-square Method (DCM) that
  can determine the best model
  for many signals superimposed 
  on arbitrary polynomial trends.
  DCM minimizes the Chi-square for 
  the data in the 
  multi-dimensional 
  tested frequency
  space. 
  The required number of tested
  frequency combinations 
  remains manageable,
  because the method
  test statistic
  is symmetric in this
  tested frequency space.
  With our known tested
  constant 
  frequency grid values, 
  the non-linear DCM model 
  becomes linear,
  and all
  results become unambiguous.
  We test DCM with
  simulated data
  containing different
  mixtures of signals
  and trends.
  DCM gives
  unambiguous results,
  if the signal frequencies
  are not too close
  to each other,
  and none of the signals
  is too weak.
  It relies on brute
  computational force,
  because all possible free
  parameter combinations for
  all reasonable linear models
  are tested.
  DCM works like winning
  a lottery
  by buying all
  lottery tickets.
  Anyone can reproduce
  all our
  results
  with the DCM computer
  code.\footnote{This python program
  \MyPro ~and the other three necessary
  files are freely available in
  \Link{https://zenodo.org/record/3661073}{Zenodo
  database: doi 10.5281/zenodo.3661072}.
  All files, variables
  and other program code
related items are 
printed in \PR{magenta} colour.
Our Appendix gives
detailed instructions for
using \MyPro.}
We also present one
  preliminary real use case,
where DCM is applied to the observed (O) minus 
the computed (C) eclipse epochs of a 
binary star, \target.
  This DCM analysis
  reveals evidence for the possible
  presence of a third and a fourth body
  in this system.
  One recent study of a very large sample
  of binary stars
  indicated that the probability for
  detecting a fourth body
  from the O-C data
  of eclipsing binaries is only about 0.00005.

\end{abstract}

\section{Introduction}
\label{SectIntroduction}

The
Discrete Fourier Transform (DFT),
also called
the power spectrum method,
is one of the most frequently
applied period analysis
methods in natural
sciences
\citep[e.g.][]{Lom76,Sca82,Zec09}.
The above DFT versions rely
on the assumption that the data
contains no trends,
and the correct model
is one sinusoidal signal.
In the Lomb-Scargle version,
  the mean of the data is removed before
  DFT computation, while
  the Zechmeister-K\"urs\-ter 
  version gives
  an unambiguous value for this mean.
  Systematic trends in the data must 
  be removed before DFT analysis,
  like in the Kepler
  satellite light curve
  detrending pipelines
  PDC-MAP \citep{Mur12}
  or
  ARC2 \citep{Aig17}.
However, this removal of trends is not
trivial, and it can seriously
mislead the period analysis
\citep[e.g.][]{Ols18}.

Since DFT searches for
one period 
at the time,
we call it 
a one-dimensional period
finding method.
All trends in the data must be
removed before applying DFT.
After this detrending,
the DFT search for many
pure sinusoidal signals
usually relies on
``pre-whitening''.
In this technique,
the highest DFT periodogram peak
gives the best period
for the detrended
{\it original}
data. 
The sinusoidal model 
with this best
period is subtracted 
from these detrended data.
The next second best period
is determined with
the DFT analysis
of the {\it residuals}.
This second best period
gives the sinusoidal model 
for the residuals, and
the next
residuals for DFT analysis.

Countless
DFT studies have been
published in
natural sciences.
Since
Astrophysics Data System
(ADS) alone contains
over four thousand
citations to the
DTF version by \citet{Sca82},
we mention only
some recent astronomical DFT studies:
Analysis of Kepler satellite
light curves \citep[e.g.][]{Rei13A},
Planet detection from radial velocities
\citep[e.g.][]{May19},
Variable star identification in
large surveys
\citep[e.g.][]{Paw19}
and 
Stellar
pulsations \citep[e.g.][]{Mel19}.

There are other period finding 
methods that can 
search for more complicated models
than a simple sinusoid,
like the Three Stage
Period Analysis
\citep[][TSPA]{Jet99} or
the Continuous Period Search
\citep[][CPS]{Leh11}.
However, these methods can also
only
detect one signal at the time.

Our DCM can detect
many signals superimposed
on arbitrary polynomial trends.
While DFT can detect
  {\it only}
  first order harmonic signals
  of pure sinusoids,
  our DCM can also detect
  much more complicated signals
  composed of any
  arbitrary order
  harmonics.
We formulate DCM
in
Sects. \ref{SectModel} and
\ref{SectMethod},
  and test it with simulated
  data in 
  Sect. \ref{SectSimulations}.
  We demonstrate how DCM
  can unambiguously detect a sum of three
  sinusoids superimposed on
  a second order polynomial trend 
  (Sect. \ref{SectSimulatedOne}),
  and how this best model
  can be identified among many 
  alternative nested models for
  the data
  (Sect. \ref{SectIdentify}).
  The consequences of searching
  for too many, or too few, signals
  are discussed (Sects. 
  \ref{SectTooMany} and
  \ref{SectTooFew}).
  Finally, we determine the 
  data constraints for an
  unambiguous DCM analysis
  (Sect. \ref{SectManyModels}).
  All these results can be 
  reproduced with the DCM
  program code
  input and 
  output 
  specified in our appendix 
  (Table \ref{TableReproduce}).
  One DCM real use case
  is also presented (Sect. \ref{SectXZAnd}).

\section{Model}
\label{SectModel}

The data are
$y_i=y(t_i) \pm \sigma_i$,
where $t_i$ are the observing
times and $\sigma_i$ are the errors
$(i=1,2, ..., n)$. 
The time span of data is $\Delta T\!=\!t_n\!-\!t_1$.
The notations for the mean and the standard deviation
of $y_i$ are $m_y$ and $s_y$.
Before modelling, we subtract the first
observing time $t_1$ from all observing times $t_i$.
Hence, the zero point in time, $t=0$, is at $t_1$.
Our model is
\begin{eqnarray}
  g(t) = g(t,K_1,K_2,K_3)
   =  h(t) + p(t), 
\label{Eqmodel}
\end{eqnarray}
where
\begin{eqnarray}
h(t)   & = & h(t,K_1,K_2) = \sum_{i=1}^{K_1} h_i(t) \label{Eqharmonicone} \\
h_i(t) & = & \sum_{j=1}^{K_2} 
B_{i,j} \cos{(2 \pi j f_i t)} + C_{i,j} \sin{(2 \pi j f_i t)}
\label{Eqharmonictwo} \\
p(t)   & = & p(t,K_3) = \sum_{k=0}^{K_3} p_k(t) \\
p_k(t) & =  & M_k \left[
                    {{2t} \over {\Delta T}}
                    \right]^k.
\label{Eqpolynomial}
\end{eqnarray}
This model searches
  for two patterns in the data:
  the periodic $h(t)$ pattern
  that repeats itself,
  and the aperiodic $p(t)$
  pattern that does not. 
  The $K_1$ harmonic $h_i(t)$
  signals have
a frequency $f_i$ and
an order $K_2$. 
The sum of these signals
is superimposed
on the $K_3$ 
order polynomial trend.
The number of free parameters is
\begin{eqnarray}
  p=K_1 \times (2K_2+1)+K_3+1.
  \label{Eqp}
\end{eqnarray}
They are 
$\bar{\beta}=$ $[\beta_1, \beta_2, ..., $
$\beta_p] = $
$[B_{1,1},C_{1,1},f_1, ..., $ 
$B_{K_1,K_2}, $ 
$C_{K_1,K_2}, f_{K_1},$ 
$M_0, ..., M_{K_3}]$.
The first group of
free parameters, the frequencies $\bar{\beta}_{I}=[f_1, ..., f_{K_1}]$,
make this $g(t)$ model non-linear.
If these $\bar{\beta}_{I}$ are fixed to constant
known numerical values, the model becomes linear,
and the solution for the 
remaining second group of free parameters,
$\bar{\beta}_{II}=[B_{1,1}C_{1,1}, ..., $
$B_{K_1,K_2}, C_{K_1,K_2},$ $M_0, ..., M_{K_3}]$,
is unambiguous.

The $2t/\Delta T$ argument
in $p_k(t)$ ensures that the scale
in the polynomial
coefficients $M_0, ..., M_{K_3}$ 
is the same as in the amplitudes
$B_{1,1}, C_{1,1}, ..., B_{K_1,K_2}, C_{K_1,K_2}$ 
of $h_i(t)$ harmonics.
With this scaling, the higher polynomial
orders can not dominate $p_k(t) >> h_i(t)$,
nor become insignificant $p_k(t) << h_i(t)$,
for any arbitrary unit of time $t$.
Therefore, the simulated values of all these
free parameters $\bar{\beta}_{II}$
can later be drawn from 
the same uniform random distribution 
(Sect. \ref{SectSimulations}: Eq. \ref{EqUniformBeta}).

The model residuals 
\begin{eqnarray}
\epsilon_i = y(t_i) - g(t_i) = y_i - g_i
\label{Eqresiduals}
\end{eqnarray}
give the Chi-square
\begin{eqnarray}
\chi^2 = \sum_{i=1}^n {{\epsilon_i^2} \over {\sigma_i^2}}
\label{EqChi}
\end{eqnarray}
and the sum of squared residuals 
\begin{eqnarray}
R= \sum_{i=1}^n \epsilon_i^2.
\label{EqR}
\end{eqnarray}

For each
$h_i(t)$ signal,
we determine the parameters 
\begin{itemize}

\item[] $P_i = 1/f_i = $ Period
\item[] $A_i = $ Peak to peak amplitude
\item[] $t_{\mathrm{i,min,1}} = $ Deeper primary minimum epoch 
\item[] $t_{\mathrm{i,min,2}} = $ Secondary minimum epoch (if present)
\item[] $t_{\mathrm{i,max,1}} = $ Higher primary maximum epoch
\item[] $t_{\mathrm{i,max,2}} = $ Secondary maximum epoch (if present)
\end{itemize}
The first observing time $t_1$,
which is removed before modelling,
is added back to the above four epochs.
The $P_i$ and $A_i$ values are the same
for any zero point $t=0$.

In our figures,
we use the same colours for
the frequencies, the amplitudes,
the curves and the
periodograms of the same $h_i(t)$ signal.
We give those colours 
in Table \ref{Tablecolour}.

\section{Method}
\label{SectMethod}

If the errors $\sigma_i$ are known,
the test statistic of our period finding method is
\begin{equation}
z=z(f_1,f_2, ..., f_{K_1})
= \sqrt{{ \chi^2 \over n }},
\label{EqZone}
\end{equation}
where $\chi^2$ is minimized for the linear $g(t)$
model having the fixed tested 
$\bar{\beta_{I}}=[f_1,f_2, ..., f_{K_1}]$ frequencies.

If the $\sigma_i$ errors are unknown,
we use 
\begin{equation}
z=z(f_1,f_2, ..., f_{K_1})= \sqrt{{
                                {R}
                                \over
                                {n}
                              }}.
\label{EqZtwo}
\end{equation}
The core of DCM approach is that
  the test statistic
  $z$ in Eqs. \ref{EqZone} and \ref{EqZtwo}
  refers indirectly to 
  the reduced Chi-square
  \citep{Bar93,And10}.
Our DCM computer code
\PR{\MyPro} minimizes $z$.
For any data,
the possible
alternative nested models 
that can be tested
with \MyPro  ~are
\begin{itemize}
\item[] $0 \le K_1 \le 6 \equiv$
  From one to six periodic signals
\item[] $1 \le K_2 \le 2 \equiv$
  Harmonic signal orders
\item[] $0 \le K_3 \le 6 \equiv$
  Polynomial trend orders.
\end{itemize}
Any arbitrary
pair, $g_1(t)$ and $g_2(t)$, of 
these nested models
can be compared.
We use
the number of free parameters $(p_1 < p_2)$, 
the Chi-squares $(\chi^2_1, \chi^2_2)$,
and 
the sum of squared residuals $(R_1,R_2)$
of these two models
to determine which one of them is
a better  model for the data.
If the errors $\sigma_i$ are known, 
our test statistic is 
\begin{eqnarray}
F_{\chi} & = &
\left(
{
{\chi^2_1}
\over
{\chi^2_2}
}
-1
\right)
\left(
{
{n-p_2-1}
\over
{p_2-p_1}
}
\right). \label{EqFChi} 
\end{eqnarray}
If these errors are unknown,
we use 
\begin{eqnarray}
F_R =
\left(
{
{R_1}
\over
{R_2}
}
-1
\right)
\left(
{
{n-p_2-1}
\over
{p_2-p_1}
}
\right).
\label{EqFR}
\end{eqnarray}
The $F_{\chi}$ or $F_R$ test statistic is used to identify
the better model for the data.
The null hypothesis is
\begin{itemize}

\item[] $H_{\mathrm{0}}$: {\it ``The model $g_2(t)$ 
does not provide
a significantly better fit to the data
than the 
model $g_{\mathrm{1}}(t)$.'' }

\end{itemize}

\noindent
Under $H_0$, both
$F_{\chi}$ and $F_R$ have an $F$ distribution with 
$(\nu_1,\nu_2)$ degrees of freedom,
where $\nu_1=p_2-p_1$ and $\nu_2=n-p_2$
\citep{Dra98}. 
The probability for 
$F=F_{\chi}$ or $F=F_R$  reaching a fixed level $F_0$
is called the critical level $Q_{F} = P(F \ge F_0)$. 
We reject the $H_0$ hypothesis, if
\begin{eqnarray}
Q_F < \gamma_F=0.001,
\label{EqQF}
\end{eqnarray}
where $\gamma_F$ is a pre-assigned significance level.

\begin{figure}
\begin{center}
  \resizebox{6.0cm}{!}
  {\includegraphics{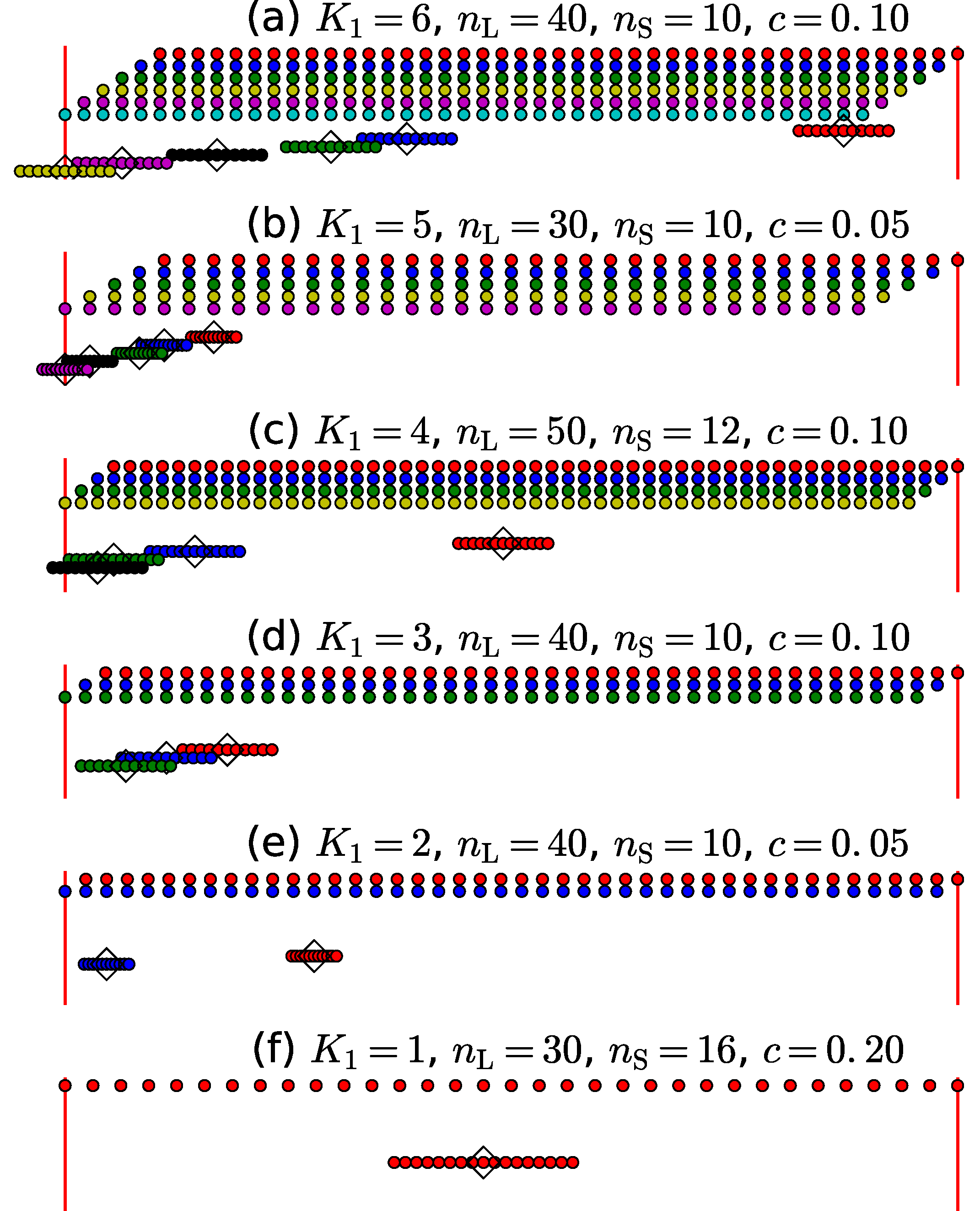}}
\end{center}
\caption{Arbitrary tested
  frequencies.
(a) Six highest longer rows show 
tested frequencies
$f_1>f_2>f_3>f_4>f_5>f_6$
for $K_1=6$ signals
and $n_{\mathrm{L}}=40$
in a {\it long} search between
$P_{\mathrm{max}}^{-1}$
and  $P_{\mathrm{min}}^{-1}$
(vertical red lines).
Six lower shorter rows show
{\it short} search frequencies
for $n_{\mathrm{S}}=10$
and $c=0.10$, where 
diamonds denote best
frequency candidates
$f_{\mathrm{i,mid}}$
detected in {\it long} search.
Symbol colours are
given in Table \ref{Tablecolour}.
(b-f) Arbitrary
tested frequencies 
for $K_1=5,4,3,2$
and 1 signals with
given $n_{\mathrm{L}}$,
$n_{\mathrm{S}}$ and
$c$ combinations.}
\label{FigGrid}
\end{figure}

For $K_1=2$ signals,
the $z(f_1,f_2)=z(f_2,f_1)$ symmetry 
requires only the testing of
$f_1>f_2$ combinations.
The six respective symmetries
 $z(f_1,f_2,f_3)=$
    $z(f_1,f_3,f_2)=$
    $z(f_2,f_1,f_3)=$
    $z(f_2,f_3,f_1)=$
    $z(f_3,f_1,f_2)=$ 
    $z(f_3,f_2,f_1)$
for $K_1=3$ signals
require only the  $f_1>f_2>f_3$
tests (e.g. Fig. \ref{FigGrid}d).
Hence, we test only the 
$f_1 > f_2 > f_3 > f_4 > f_5 > f_6$
combinations.

\begin{figure}
\begin{center}
  \resizebox{8.5cm}{!}
  {\includegraphics{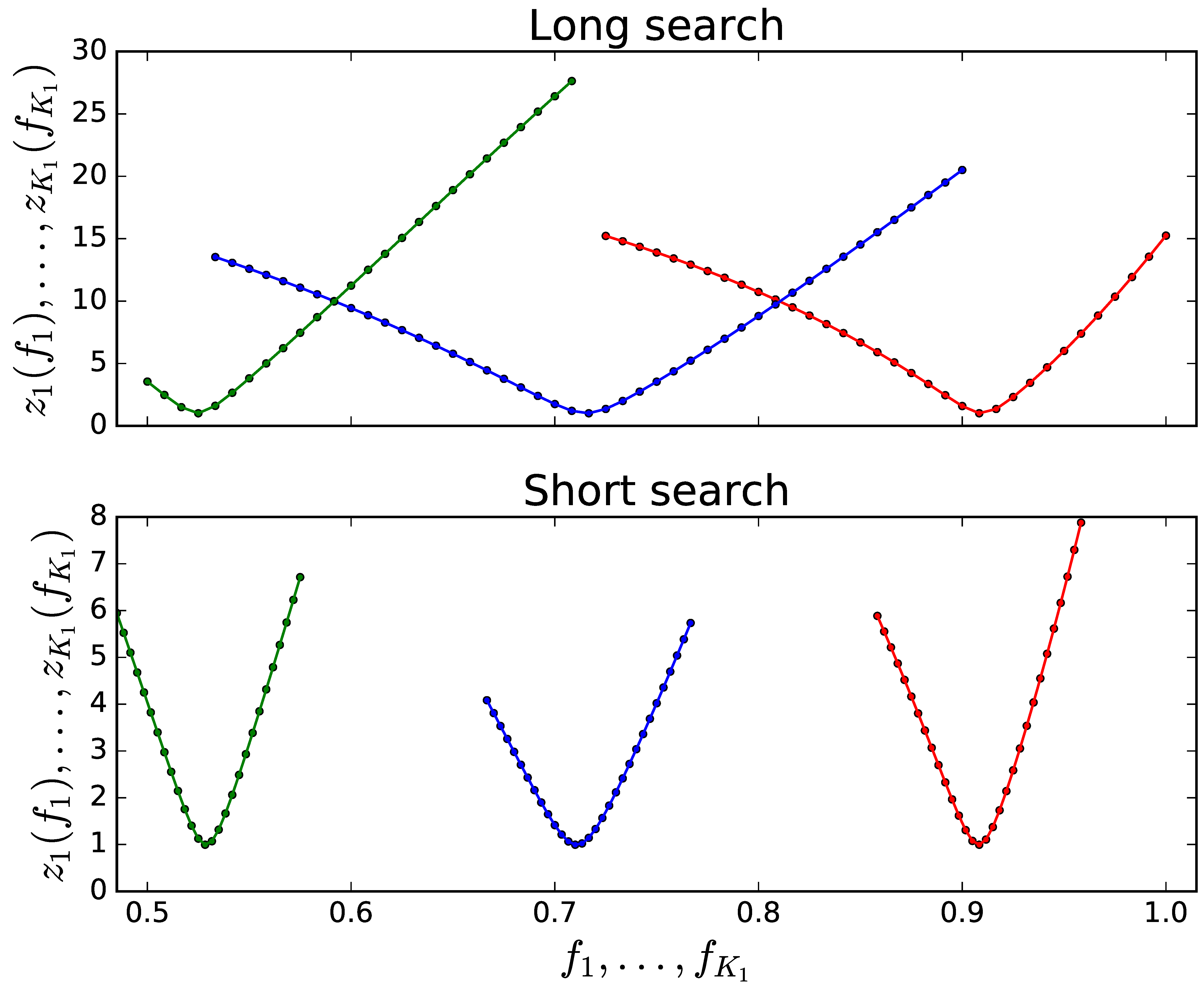}}
\end{center}
\caption{Model 19 long and
  short search
  periodograms for 
  Table \ref{TableSimulatedData}
  data.
  Colours for
  $z_1(f_1)$, $z_2(f_2)$
  and $z_3(f_3)$ 
  are given in
  Table \ref{Tablecolour}.}
\label{FigZ}
\end{figure}

In our long frequency
interval search,
we test an evenly
spaced long grid of $n_{\mathrm{L}}$ 
frequencies between
$f_{\mathrm{min}}=P_{\mathrm{max}}^{-1}$
and $f_{\mathrm{max}}=P_{\mathrm{min}}^{-1}$
(Figs. \ref{FigGrid}a-f:
higher longer rows). 
The best frequency candidates 
$f_{\mathrm{1,mid}}, ...,
f_{\mathrm{K_1,mid}}$ 
at the $z$ minimum give 
the mid points for
the denser evenly spaced 
short grids
of $n_{\mathrm{S}}$ tested frequencies
(Fig. \ref{FigGrid}: diamonds).
The intervals of these
short grids are
\begin{eqnarray}
  [f_{\mathrm{i,mid}}-a,
  f_{\mathrm{i,mid}}+a].
  \label{EqShort}
\end{eqnarray}
The suitable values are
$a =
c~(f_{\mathrm{max}}-f_{\mathrm{min}})/2$,
where the width is
$5\% \equiv 0.05 \le c
\le 0.20 \equiv 20 \%$ 
of the long test interval
(Figs. \ref{FigGrid}a-f:
lower shorter rows). 
The best frequencies are
at the global minimum
of the periodogram 
\begin{eqnarray}
  z_{\mathrm{min}}=
  z(f_{\mathrm{1,best}},f_{\mathrm{2,best}},...,
  f_{\mathrm{K_1,best}}).
\end{eqnarray}
Some graphical presentation
of the full $z$
periodogram
would be possible only
for the one $z(f_1)$,
the two  $z(f_1,f_2)$
and 
the three  $z(f_1,f_2,f_3)$
dimensional cases.
We solve these dimensional
problems by
presenting only the
following one-dimensional
slices of the full periodograms
\begin{eqnarray}
  z_1(f_1) & = & z(f_1,f_{\mathrm{2,best}}, ...,f_{\mathrm{K_1,best}}) \nonumber \\
  z_2(f_2) & = & z(f_{\mathrm{1,best}},f_2, f_{\mathrm{3,best}}, ...,f_{\mathrm{K_1,best}}) \nonumber \\
  z_3(f_3) & = & z(f_{\mathrm{1,best}},f_{\mathrm{2,best}},f_3, f_{\mathrm{4,best}},...,f_{\mathrm{K_1,best}}) ~\label{EqSlices} \\
  z_4(f_4) & = & z(f_{\mathrm{1,best}},f_{\mathrm{2,best}},f_{\mathrm{3,best}},f_4, f_{\mathrm{5,best}},f_{\mathrm{K_1,best}}) \nonumber \\
  z_5(f_5) & = & z(f_{\mathrm{1,best}},f_{\mathrm{2,best}},f_{\mathrm{3,best}},f_{\mathrm{4,best}},f_5,f_{\mathrm{K_1,best}}) \nonumber \\
  z_6(f_6) & = & z(f_{\mathrm{1,best}},f_{\mathrm{2,best}},f_{\mathrm{3,best}},f_{\mathrm{4,best}},f_{\mathrm{5,best}},f_6) \nonumber 
\end{eqnarray}
All best frequencies fulfill 
$f_{\mathrm{i,best}}\! >\!
f_{\mathrm{i+1,best}}$,
because we test only frequencies
$f_1\!>\!f_2\!>\!f_3\!>\!f_4\!
>\!f_5\!>\!f_6$.
Therefore,
every $z_{i}(f_{i})$ periodogram
ends at the
minimum of
the next $z_{i+1}(f_{i+1})$
periodogram
(e.g. Fig. \ref{FigZ}: upper panel).

We perform a {\it linear}
least squares fit to the
data with the fixed
numerical values of the
best frequencies 
$\bar{\beta}_{\mathrm{I,Initial}}=
[f_{\mathrm{1,best}},f_{\mathrm{2,best}},...,
f_{\mathrm{K_1,best}}]$
detected in the short
interval search.
This gives us the 
unambiguous estimates
for the values of
the other free parameters
$\bar{\beta}_{\mathrm{II,Initial}}$.
We determine the final estimates
for the free parameters
with the
standard {\it non-linear}
least squares iteration
\begin{eqnarray}
  \bar{\beta}_{\mathrm{Initial}}
  \rightarrow
  \bar{\beta}_{\mathrm{Final}},
\label{EqInitial}
\end{eqnarray}
where $\bar{\beta}_{\mathrm{Initial}}=
[\bar{\beta}_{\mathrm{I,Initial}},
\bar{\beta}_{\mathrm{II,Initial}}]$.

The errors for
the model parameters are
determined with
the bootstrap procedure
\citep{Efr86,Jet99}.
For the 
{\it original data},
we test all frequency combinations 
within the short intervals
of Eq. \ref{EqShort}.
During each bootstrap round,
we select
a random sample $\bar{\epsilon}^*$
from
the residuals $\bar{\epsilon}$
of this best $g(t)$ model
for the {\it original}
data $\bar{y}$ 
(Eq. \ref{Eqresiduals}).
Any $\epsilon_i$ value
can enter into this random
sample $\bar{\epsilon}^*$
as many times as
the random selection 
happens to favour it.
These random residuals
give the {\it artificial} 
data sample
\begin{eqnarray}
y_i^*=g_i+\epsilon_i^*
\label{EqBoot}
\end{eqnarray} 
during each bootstrap round.
The best model
for each artificial
$\bar{y}^*$ random
data sample gives
one estimate for
every model parameter.
The error estimate
for each particular 
model parameter is
the standard deviation
of all estimates obtained
for this parameter 
in all bootstrap rounds.

\begin{figure*}
\begin{center}
  \resizebox{17.0cm}{!}
  {\includegraphics{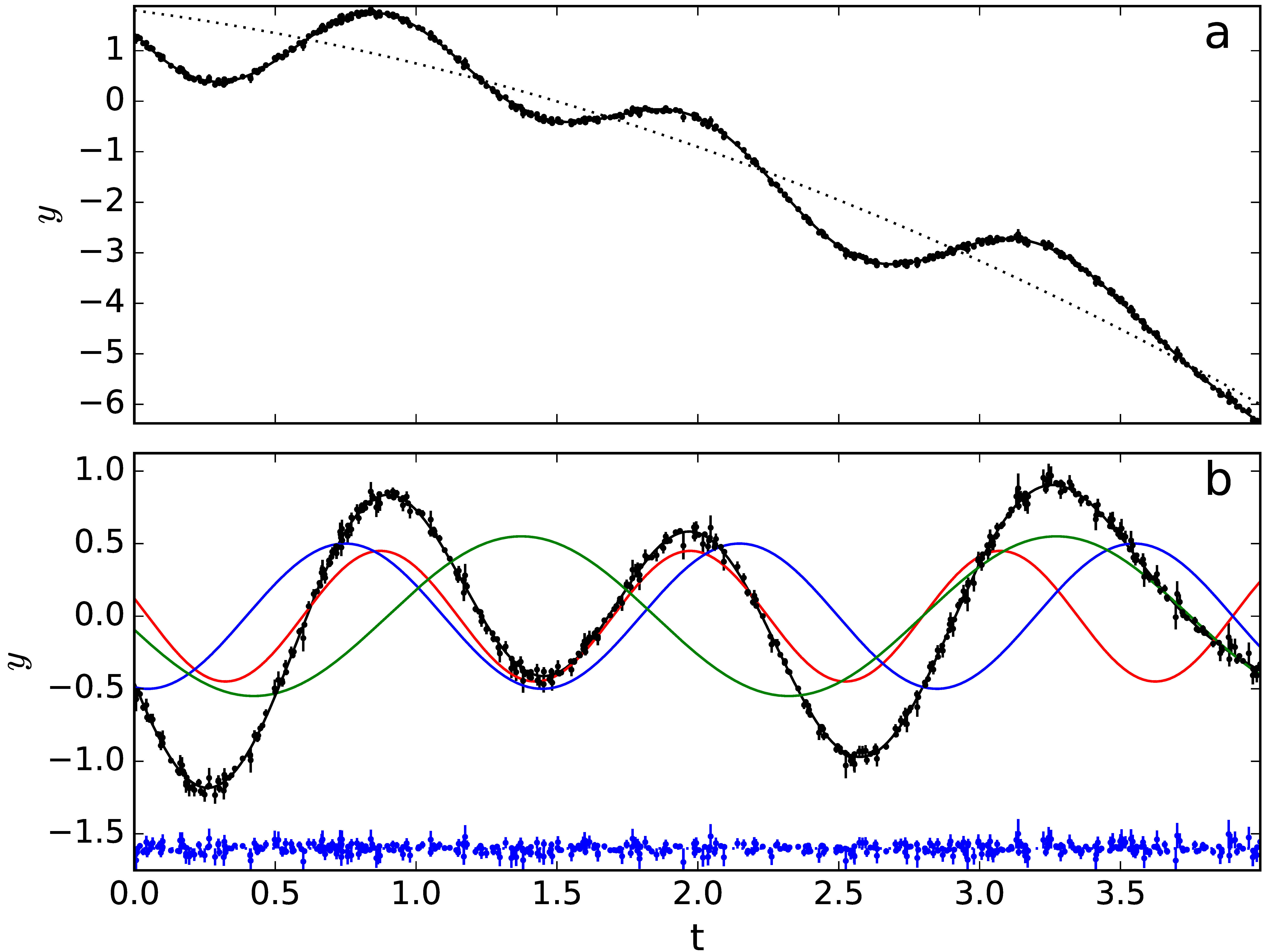}}
\end{center}
\caption{Model 19 for  Table
  \ref{TableSimulatedData} data.
  (a) Data $y_i \pm \sigma_1$
  (black circles),
  and $g(t)$ and $p(t)$ curves.
  (b) Data minus $p(t)$,
  $h(t)$, $h_1(t)$,
  $h_2(t)$ and
  $h_3(t)$ curves.
  Residuals (blue circles)
  are offset to $y=-1.6$ level.
  Colours of all curves
  are given in
  Table \ref{Tablecolour}
}
\label{FigModel}
\end{figure*}

\begin{figure}
\begin{center}
  \resizebox{8.5cm}{!}
   {\includegraphics{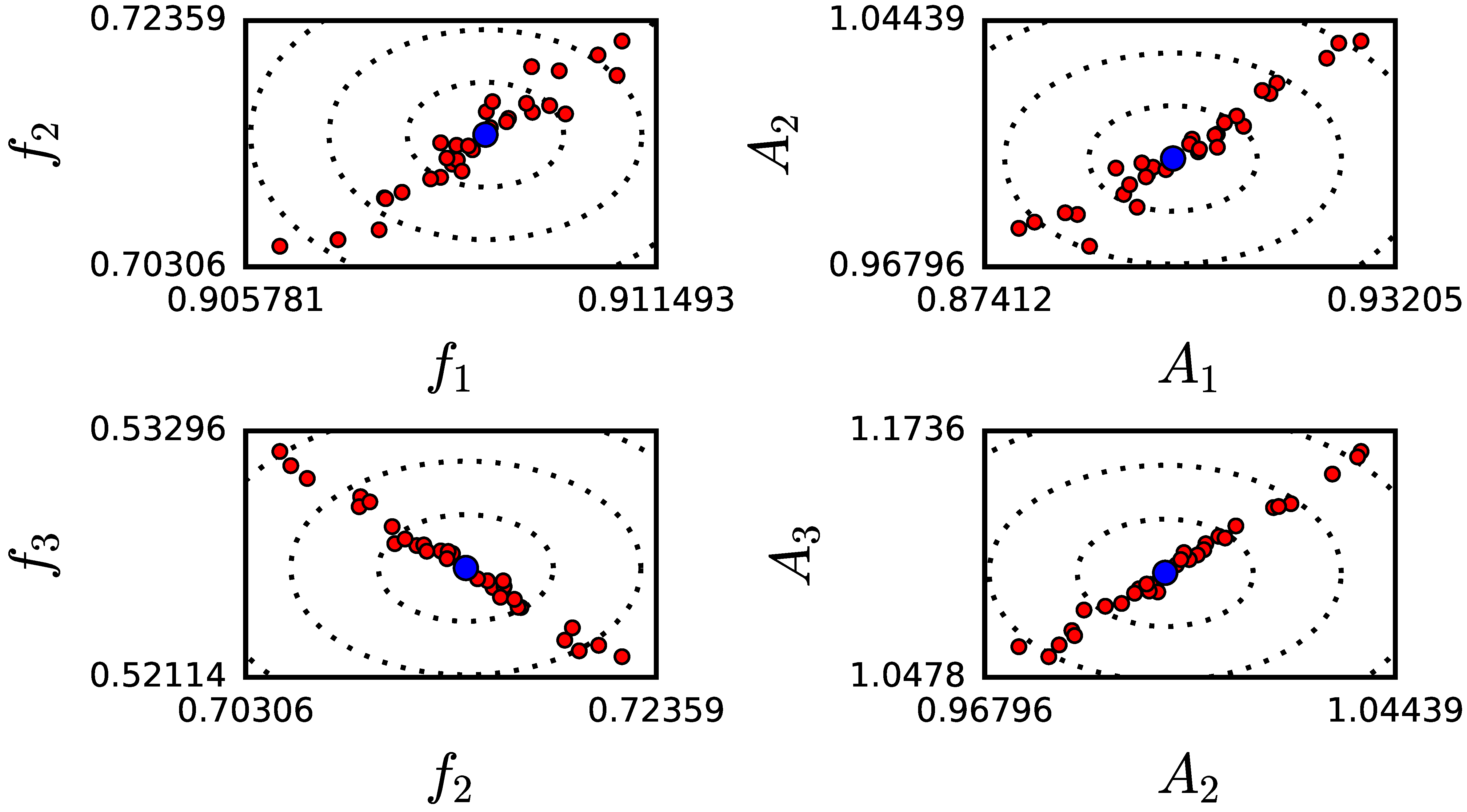}}
\end{center}
\caption{Model 19
  frequencies
  ($f_1,f_2, f_3$) and
  amplitudes ($A_1,A_2,A_3$).
  Original model results 
  are denoted with
  larger blue circles
  and bootstrap
  results with smaller
  red circles.
  Dotted lines denote
  one, two and three sigma
  error limits.}
\label{FigTwoSample}
\end{figure}

\section{Simulated data}
 \label{SectSimulations}

 We test our method
 with simulated data.

\subsection{One simulated model}
\label{SectSimulatedOne}

Here, we show that
our method can
detect the correct 
model parameter values.
We illustrate this
with the following model
having known free parameter values
\begin{eqnarray}
g_{\mathrm{S1}}(t) & = & h(t) + p(t) 
= \sum_{i=1}^{K_1} h_i(t) +  \sum_{k=0}^{K_3} p_k(t)
\label{Eqsimulationone} \\
  h_i  & = &
(A_i/2) \sin{[2 \pi f_i (t-T_i)]}
             \nonumber \\
p_k(t) & =  & M_k \left[
{{2t} \over {\Delta T}}
               \right]^k, \nonumber
\end{eqnarray}
where
$K_1\!=\!3$ and $K_3\!=\!2$.
The order of the three sinusoidal
 $h_i(t)$  signals is $K_2=1$.
 The adopted known free
 parameter values
   $f_1\!=\!1/P_1$,
   $f_2\!=\!1/P_2$,
   $f_3\!=\!1/P_3$,
    $T_1$,
    $T_2$,
    $T_3$,
    $A_1$,
    $A_2$,
    $A_3$,
    $M_0$,
    $M_1$ and
    $M_2$
    are given in Table
    \ref{TableSimOne}.

    The simulated $n^{\star}=500$ time
points $t_i^{\star}$ are
drawn from a uniform
random distribution
\begin{eqnarray}
  U(0,\Delta T,n^{\star})
  \label{EqRandomT}
\end{eqnarray}
between 0 and $\Delta T=4$.

\tablepage
\begin{table*}
  \caption[]{Figure notations
    and colours.}
\begin{center}
\begin{tabular}{lclllccllccll}
\hline
  & \multicolumn{3}{c}{Frequencies, Amplitudes}
  & & \multicolumn{3}{c}{Functions}
  & & \multicolumn{3}{c}{Periodograms} \\
\cline{2-5} \cline{7-9} \cline{11-13}
Colour    & $i$ & $f_i,A_i$  & Symbol  & Figs. & 
& $h_i(t)$ & Symbol                & Figs.       
& & $z_i(t)$   & Symbol            & Figs.    \\
\hline
Black        &  -  &   -    &  -   &        -
& & $g(t)$   & Continuous line     &  \ref{FigModel}a, \ref{FigTooManyTwo}a, \ref{FigTooFewTwo}a
&         & -   & -                &        -                \\
Black     & -   & -      & -       &     - 
& & $p(t)$   & Dotted line         &  \ref{FigModel}a, \ref{FigTooManyTwo}a, \ref{FigTooFewTwo}a
& & -         & -                  &     -               \\
Black     & -   & -      & -       &     - 
& & $h(t)$   & Continuous line     &  \ref{FigModel}b, \ref{FigTooManyTwo}b, \ref{FigTooFewTwo}b
& & -         & -                  &     -               \\
Red   & 1   & $f_1,A_1$  & Circles &  \ref{FigGrid}, \ref{FigTwoSample}, \ref{FigTooManyThree}, \ref{FigTooFewThree}, \ref{FigMany}  
& & $h_1(t)$ & Continuous line     &  \ref{FigModel}b, \ref{FigTooManyTwo}b, \ref{FigTooFewTwo}b
& & $z_1(f_1)$ & Continuous line   &    \ref{FigZ},  \ref{FigTooManyOne},  \ref{FigTooFewOne}        \\
Blue  & 2   & $f_2,A_2$  & Circles &   \ref{FigGrid}, \ref{FigTwoSample}, \ref{FigTooManyThree}, \ref{FigTooFewThree}, \ref{FigMany} 
& & $h_2(t)$ & Continuous line     &  \ref{FigModel}b, \ref{FigTooManyTwo}b, \ref{FigTooFewTwo}b
& & $z_2(f_4)$ & Continuous line   &    \ref{FigZ},  \ref{FigTooManyOne},  \ref{FigTooFewOne}        \\
Green  & 3   & $f_3,A_3$  & Circles &  \ref{FigGrid}, \ref{FigTwoSample}, \ref{FigTooManyThree}, \ref{FigMany} 
& & $h_3(t)$ & Continuous line     & \ref{FigModel}b, \ref{FigTooManyTwo}b
&  & $z_3(f_3)$ & Continuous line  &   \ref{FigZ},  \ref{FigTooManyOne}         \\
Yellow & 4   & $f_4,A_4$  & Circles&  \ref{FigGrid}, \ref{FigTooManyThree}
& & $h_4(t)$ & Continuous line     &  \ref{FigTooManyTwo}b           
& & $z_4(f_4)$ & Continuous line   &  \ref{FigTooManyOne}          \\
Magenta& 5   & $A_5,f_5$  & Circles &  \ref{FigGrid}        
& & $h_5(t)$ & Continuous line     &        -     
& & $z_5(f_5)$ & Continuous line   &        -          \\
Cyan  & 6   & $f_6,A_6$  & Circles  &  \ref{FigGrid}        
& & $h_6(t)$ & Continuous line     &        -
& & $z_6(f_6)$ & Continuous line   &        -           \\
\hline
\end{tabular}
\end{center}
\label{Tablecolour}
\end{table*}

Evenly spaced observations
$y_i$ coinciding with
the sinusoid
$g_i=$ $a \sin{2 \pi t_i}$
fulfill $a= 2^{3/2} s_y$,
where $s_y$ is
the standard deviation of $y_i=g_i$
\citep[][]{Jet13}. 
The peak to peak amplitude
of such a sinusoid fulfills
$A = 2 a = 2^{5/2} s_y$.
This relation also holds
for cosine, double sine
and double cosine curves.
Therefore, we compute
an estimate for
the peak to peak amplitude $A$
of the simulated
periodic signal from
the standard deviation
$s_y^{\star}$ of
the sum $h(t^{\star}_i)$ of all
$h_i(t^{\star}_i)$ signals
in Eq. \ref{Eqsimulationone}.
If $\sigma_m$ is
the mean of
all data errors $\sigma_i$,
the signal to noise ratio 
$\mathrm{SN} =
A/\sigma_m=
2^{5/2} s_y^{\star} /\sigma_m$
gives
\begin{equation}
  \sigma_m =
  2^{5/2} s_y^{\star}/ \mathrm{SN}
\label{EqSigmaSim}
\end{equation}
for the accuracy
of simulated data.
For our chosen
fixed ${\mathrm{SN}}=100$ level, 
we draw the simulated
data errors
$\sigma_i^{\star}$ 
from a Gaussian distribution
\begin{eqnarray}
N(m^{\star},s^{\star},n^{\star}), 
\label{EqError}
\end{eqnarray}
where $m^{\star}=0$
and $s^{\star}=\sigma_m$.
The numerical values
for one arbitrary
sample of simulated
data\footnote{Simulated
  Table
  \ref{TableSimulatedData}
  data are in
  file \MyTest ~in Zenodo.}
\begin{eqnarray}
  y_i^{\star}=
  g_{\mathrm{S1}}(t^{\star})
  +\sigma_i^{\star}
\label{EqSimulatedData}
\end{eqnarray}
are given in Table
\ref{TableSimulatedData}.

We perform the period analysis
for the simulated data 
of Table \ref{TableSimulatedData}
over
the long tested
period interval  between 
$P_{\mathrm{min}}=1$
and $P_{\mathrm{max}}=2$.
The test statistic $z$
of Eq. \ref{EqZone} is computed
for the $g(t,K_1=3,K_2=1,K_3=2)$
model of Eq. \ref{Eqmodel},
which will be later
referred to as ``model 19''
(see Table \ref{TableModels}).
The $z_1(f_1)$, $z_2(f_2)$
and $z_3(f_3)$
periodograms 
are shown in Fig. \ref{FigZ},
where all three minima
are clearly separated.
The simulated input and
the detected
output model parameter
values
are given in
Table \ref{TableSimOne}.
They agree perfectly.
This best detected model 19
for the simulated data is
shown in Fig. \ref{FigModel}a.
The residuals are stable
and show no systematic trends
(Fig. \ref{FigModel}b: blue circles).
The results for
the frequencies ($f_1,f_2,f_3$)
and
the amplitudes ($A_1,A_2,A_3$)
of this best model
are shown in
Fig. \ref{FigTwoSample}.
The bootstrap estimates
for the
frequencies and the
amplitudes show
linear correlations.
These linear correlations indicate
that any shift away
from the correct model
in one frequency or amplitude
is compensated by a shift
in all other frequencies
and amplitudes.

\subsection{Identifying the best
  model}
\label{SectIdentify}

Here, we show {\it how}
the best model for the data
can be identified among
the many alternative
nested models. 
In the previous
Sect. \ref{SectSimulatedOne},
we simulated the data 
of Table \ref{TableSimulatedData}
with 
a model 
having $K_1=3$, $K_2=1$
and $K_3=2$
(Eq. \ref{Eqsimulationone}).
  Since we knew that this model
  was used in creating
  these data,
  we used  
  the test statistic
  $z$ for the $g(t,3,1,2)$ model
  in our period analysis.
  If the simulated data
  were real data,
  we would not necessarily
  know this correct
  $K_1=3$, $K_2=1$
  and $K_3=2$ combination. 

Let us assume that
the data of
Table \ref{TableSimulatedData}
were real data,
and the correct
$K_1, K_2$ and $K_3$
combination
would be unknown.
In that case,
we would have to test
numerous alternative models.
Therefore, we test all
$1 \le K_1 \le 4$,
$1 \le K_2 \le 2$ and
$0 \le K_3 \le 3$
combinations for
the data of
Table \ref{TableSimulatedData}.
These
32 alternative models
are compared
in Table \ref{TableModels},
where we compute
the values for
their statistical
parameters $\chi^2$,
$F_{\chi}$
and
$Q_F$. 
We compare
the ``correct model 19'' 
to all other
31 alternative models.

\begin{table}
  \caption{Simulated model.
    Column 1
    gives parameters of 
Eq. \ref{Eqsimulationone}.
Column 2 gives
simulated values.
Columns 3 and 4 give detected values
for models 19 and 20. }
\begin{center}
\addtolength{\tabcolsep}{-0.05cm}
  \begin{tabular}{ccccccc}
    \hline
                    &     & Simulated  & & \multicolumn{3}{c}{Detected} \\
\cline{3-3} \cline{5-7}
Parameter           &     & $g_{S1}(t)$ & & Model 19           & & Model 20 \\
    \hline
  $1/f_1=P_1$       &      & 1.1        & & $1.100 \pm 0.001$   & &$1.104 \pm 0.003$ \\
  $A_1$             &      & 0.9        & & $0.90  \pm 0.01$    & &$0.95  \pm 0.03$  \\
  $t_{\mathrm{1,min}}$  &      & 0.325      & & $0.325 \pm 0.001$   &  &$0.322 \pm 0.002$ \\
  $t_{\mathrm{1,max}}$  &      & 0.875      & & $0.875 \pm 0.001$   & & $0.874 \pm 0.001$ \\
  $1/f_2=P_2$       &      & 1.4       &  & $1.40  \pm 0.01$    & & $1.50 \pm 0.04$   \\
  $A_2$             &      & 1.0       &  & $1.00 \pm 0.02$     & & $1.8 \pm 0.3$ \\
  $t_{\mathrm{2,min}}$ &       & 0.050     &  & $0.50  \pm 0.01$    & & $1.441\pm 0.003 $\\
 $t_{\mathrm{2,max}}$ &        & 0.75      &  & $0.750 0\pm 0.006$  & & $0.69 \pm 0.02 $\\
  $1/f_3=P_3$       &       & 1.9      &  & $1.90 \pm 0.01$     & &$1.73 \pm 0.07 $ \\
  $A_3$             &       & 1.1      &  & $1.10   \pm 0.03$   & & $1.94 \pm 0.3 $\\
  $t_{\mathrm{3,min}}$ &       & 0.425    &   & $0.426 \pm 0.008$   & & $0.54 \pm 0.05$ \\
  $t_{\mathrm{3,max}}$ &       & 1.375    &   & $1.375 \pm 0.002$   & & $1.41 \pm 0.01$\\
  $M_0$            &       & 1.8      &   & $1.800 \pm 0.002$   & & $1.74\pm 0.02$ \\
  $M_1$            &       & -1.5     &   & $-1.500 \pm 0.003$  & & $-1.1 \pm 0.2$  \\
  $M_2$            &       & -1.2     &   & $-1.201 \pm 0.001$  & & $-1.8 \pm 0.2$ \\
  $M_3$            &       & -        &   & -                   & & $0.21  \pm 0.08$ \\
   \hline
\end{tabular}
\addtolength{\tabcolsep}{+0.05cm}
\end{center}
\label{TableSimOne}
\end{table}

This correct model 19 has
$p_2\!=\!12$ free parameters.
The fifteen alternative
models 1-13 and  17-18 have
$p_1\!<\!p_2\!=\!12$. 
The critical levels
for all these fifteen alternative
models are so low that they
fall below the
computational\footnote{We refer
  to the computational accuracy of
  \PR{f.cdf} subroutine in
\PR{scipy.optimize} python library.}
accuracy of $10^{-16}$
(Table \ref{TableModels}:
$Q_F<10^{-16}$).
\kista{Hence, any
  correct model must have at least
  $p=12$ free parameters, and}
we have to reject the better
model $H_0$ hypothesis 
presented in Sect. \ref{SectMethod}.
The correct model 19 is 
certainly better
than any of
these fifteen alternative models.

\begin{table}
  \caption{Simulated
    data\footnote{
      This data file \MyTest ~is 
    given in Zenodo.}
    of
    Eq. \ref{EqSimulatedData}.
    Only
    first two observations
    of all $n^{\star}=500$
    observations
    are shown below.}
  \begin{center}
  \begin{tabular}{ccc}
    $t_i^{\star}$ & $y_i^{\star}$       & $\sigma_i^{\star}$ \\
    0.001954782 &    1.285584396    &  0.053913136      \\
    0.008301549 &    1.222326656    &  0.082413098      \\
    ...         & ...               & ...               \\
  \end{tabular}
\end{center}
\label{TableSimulatedData}
\end{table}

We give no $F_{\chi}$ or $Q_F$ 
estimate
(Eqs. \ref{EqFChi} and \ref{EqQF})
for model 14, because it
has the same number
of free parameters
as the correct model 19.
However, model 19 is definitely
better, because its  $\chi^2=496.10$ 
is smaller
than the $\chi^2=750.28$ for
the alternative model 14
(Table \ref{TableModels}).

All remaining \kista{fifteen}
alternative
models have more
free parameters
than the correct model 19.
Hence, the number
of free parameters
for this model 19 becomes 
$p_1=12$ in Eq. \ref{EqFChi},
while that for
the other models becomes $p_2$.

The correct model 19
is better than the
following \kista{four} 
alternative models
\kista{15, 21, 25 and 26},
because they all
have higher $\chi^2$ values 
(Table \ref{TableModels}).
We refer to these
  \kista{four}
  models as
  $F_{\chi}<0$, and use $Q_F=1$ for
  their critical level,
  because there is
  certainly no reason
  to
  reject
  the better model $H_0$ hypothesis 
  presented
  in Sect. \ref{SectMethod}.

The remaining \kista{eleven}
alternative 
models \kista{16, 20, 22-24
  and 27-32}
have lower $\chi^2$
values than
the correct model 19.
However, their
critical levels $Q_F$
are far above
$\gamma_F=0.001$
(Eq. \ref{EqQF}).
This means that the better
model $H_0$ hypothesis 
is not rejected, and
the correct model 19
is also better than 
all these \kista{eleven}
alternative models.

In general,
\kista{the $\chi^2=496.10$
  value}
for model 19
is already so close
to $n-p_1=500-12=488$
that it is statistically 
impossible to reach
significant high $F_{\chi}$
values 
with more complex
$p_2 > p_1$ models, because 
increasing $p_2$
can not actually decrease
$\chi^2$ a lot.

The second best 
model for the data
is model 20 reaching $Q_F=0.078$
(Table \ref{TableModels}). 
It
resembles
the correct model 19.
Except for the third order 
polynomial coefficient
$M_3$, the other 
free parameters of this
model 20 are exactly the same  
as those of the correct model 19
(Table \ref{TableSimOne}).
As explained in Sect.
\ref{SectModel},
the scale of all polynomial
trend $p(t)$ coefficients $M_1$,
$M_2$ and $M_3$ is the same,
which means that equal
absolute values for
these coefficients
 cause the
same $p(t)$ change during $\Delta T$.
The \kista{
$|M_3|=0.21$ coefficient of 
model 20 is much smaller
than the $|M_1|=1.1$ and $|M_2|=1.8$
coefficients,} which 
means that the
first and second order 
trends dominate
over the third order trend. 
\kista{The $P_1=1.104 \pm 0.003$ 
period of model 20 
agrees with the simulated 
$P_1=1.1$ period
of the $g_{\mathrm{S1}}(t)$ model,}
but the results 
for the $P_2$ 
and $P_3$ periods 
do not. 
These results confirm 
that even a minor
deviation away from 
the correct $p(t)$
trend can mislead 
the period analysis.

We conclude that
the best model 19 for the data
can be unambiguously
identified among all
alternative 32 nested models.

\begin{table}
  \caption{Comparing 32
    nested models.
    Parameters are
    $K_1$, $K_2$, $K_3$
    (Eq. \ref{Eqmodel}),
    $p$ (Eq. \ref{Eqp}),
    $\chi^2$ (Eq. \ref{EqChi}), 
     $F_{\chi}$ (Eq. \ref{EqFChi})
     and
     $Q_F$ (Eq. \ref{EqQF}).
     Critical levels $Q_F$
     are computed for Model 19.
     Last column 
     gives failed models 
     (Fail=Yes or No)}
\begin{center}
\addtolength{\tabcolsep}{-0.08cm}
\begin{tabular}{ccccccccc}
  \hline
Model  & $K_1$   & $K_2$ & $K_3$ & $p$ & $\chi^2$  & $F_{\chi}$ & $Q_F$  & Fail       \\
  \hline
   &       &       &       &     &                     &                     &               &   \\
\multicolumn{9}{c}{One signal}  \\ \cline{2-9}
 1 &  1    &  1    &  0    &  4  & $1.12 \times 10^8$  & $1.37 \times  10^7$  &  $<10^{-16}$   & -  \\
%
%
 2 &  1    &  1    &  1    &  5  & $3.50 \times 10^6$  & $4.91 \times 10^5 $ &  $<10^{-16}$   & -  \\
 3 &  1    &  1    &  2    &  6  & $2.28 \times 10^6$  & $3.73 \times 10^5 $ &  $<10^{-16}$   & -  \\
 4 &  1    &  1    &  3    &  7  & $7.35 \times 10^5$  & $1.44 \times 10^5 $ &  $<10^{-16}$   & -  \\
 5 &  1    &  2    &  0    &  6  & $9.32 \times 10^7$  & $1.53 \times 10^7 $ &  $<10^{-16}$   & -  \\
 6 &  1    &  2    &  1    &  7  & $3.41 \times 10^6$  & $6.69 \times 10^5 $ &  $<10^{-16}$   & -  \\
 7 &  1    &  2    &  2    &  8  & $2.21 \times 10^6$  & $5.42 \times 10^5 $ &  $<10^{-16}$   & -  \\ 
 8 &  1    &  2    &  3    &  9  & $7.10 \times 10^5$  & $2.32 \times 10^5 $ &  $<10^{-16}$   & -  \\
   &       &       &       &     &                     &                     &              &     \\
  \multicolumn{9}{c}{Two signals}  \\ \cline{2-9}
 9 &  2    &  1    &  0    &  7  & $7.98 \times 10^7$  & $1.57 \times 10^7 $ &  $<10^{-16}$   & Yes\\  
%
%
10 &  2    &  1    &  1    &  8  & $1.87 \times 10^6$  & $4.59 \times 10^5$  &  $<10^{-16}$   &  No\\
%
%
11 &  2    &  1    &  2    &  9  & $6.94 \times 10^4$  & $2.25 \times 10^4$  &  $<10^{-16}$   &  No\\
%
%
12 &  2    &  1    &  3    & 10  & $2.02 \times 10^4$  & $1.71 \times 10^4$  &  $<10^{-16}$   & No \\
%
%
13 &  2    &  2    &  0    & 11  & $7.38 \times 10^7$  & $7.24 \times 10^7$ &  $<10^{-16}$   & Yes \\  
%
%
14 &  2    &  2    &  1    & 12  & $750.28$  &  -                  &  -           & \kista{No} \\
%
%
15 &  2    &  2    &  2    & 13  & $1012             $  & $<0$              &     1         & No \\
%
%
16 &  2    &  2    &  3    & 14  &  495.76              & 0.116              &  0.847        & No \\
%
%
   &       &       &       &     &                     &                     &              &   \\
\multicolumn{9}{c}{Three signals}  \\ \cline{2-9}
17 &  3    &  1    &  0    & 10  & $1.02 \times 10^7$  & $5.01 \times 10^6$  & $<10^{-16}$    & Yes \\
%
%
  18 &  3    &  1    &  1    & 11  & $1.05 \times 10^5$  &  $1.03 \times 10^5$ &  $<10^{-16}$  & Yes \\
%
%
19 &  3    &  1    &  2    & 12  &  496.10               & -                  &  -          &  No \\
%
%
20 &  3    &  1    &  3    & 13  &  492.94                & 3.116              & 0.078      &  No  \\
%
%
21 &  3    &  2    &  0    & 16  & 1290                  & $<0$              & 1           & Yes \\  
%
%
22 &  3    &  2    &  1    & 17  & 492.31               &  0.742            &  0.592     & Yes \\
%
%
23 &  3    &  2    &  2    & 18  &    487.46              & 1.421          & 0.205      &  No \\
%
%
24 &  3    &  2    &  3    & 19  &  486.75                & 1.317         & 0.240     & Yes  \\
%
%
   &       &       &       &     &                        &                &           &    \\
\multicolumn{9}{c}{Four signals}  \\ \cline{2-9}
25  & 4    & 1      & 0    & 13  &  $7.22 \times 10^5$    &  $<0$          &    1      & Yes \\
%
%
26 & 4      & 1      & 1    & 14   &  1091.33              &  $<0$         &    1      &  Yes   \\
%
%
27 & 4      & 1      & 2    & 15    &   490.95             &  1.692       & 0.168      &  Yes   \\
%
%
28 & 4      & 1      & 3    &  16  &    490.19              & 1.455       & 0.214     & Yes  \\
%
%
29 & 4      & 2      & 0   &  21   &   486.61              & 1.036        &  0.410    & Yes \\
%
%
30 & 4      & 2      & 1   &  22   &   486.62               &  0.929        & 0.506     & Yes    \\
%
%
31 & 4      & 2      & 2   &  23    &  486.60               & 0.844        & 0.595      & Yes   \\
%
%
32 & 4      & 2      & 3   & 24    &  486.10                & 0.814       &  0.636     & Yes    \\
  \hline
%
%
%
%
\addtolength{\tabcolsep}{+0.08cm}
\end{tabular}
\end{center}
\label{TableModels}
\end{table}

\subsection{Searching for
  too many signals}
\label{SectTooMany}

The simulated data of Table 
\ref{TableSimulatedData} contains 
only three signals.
Here, we check what happens,
if four signals
(i.e. too many signals)
are searched for
in these data.

\begin{figure}
\begin{center}
\resizebox{8.5cm}{!}
{\includegraphics{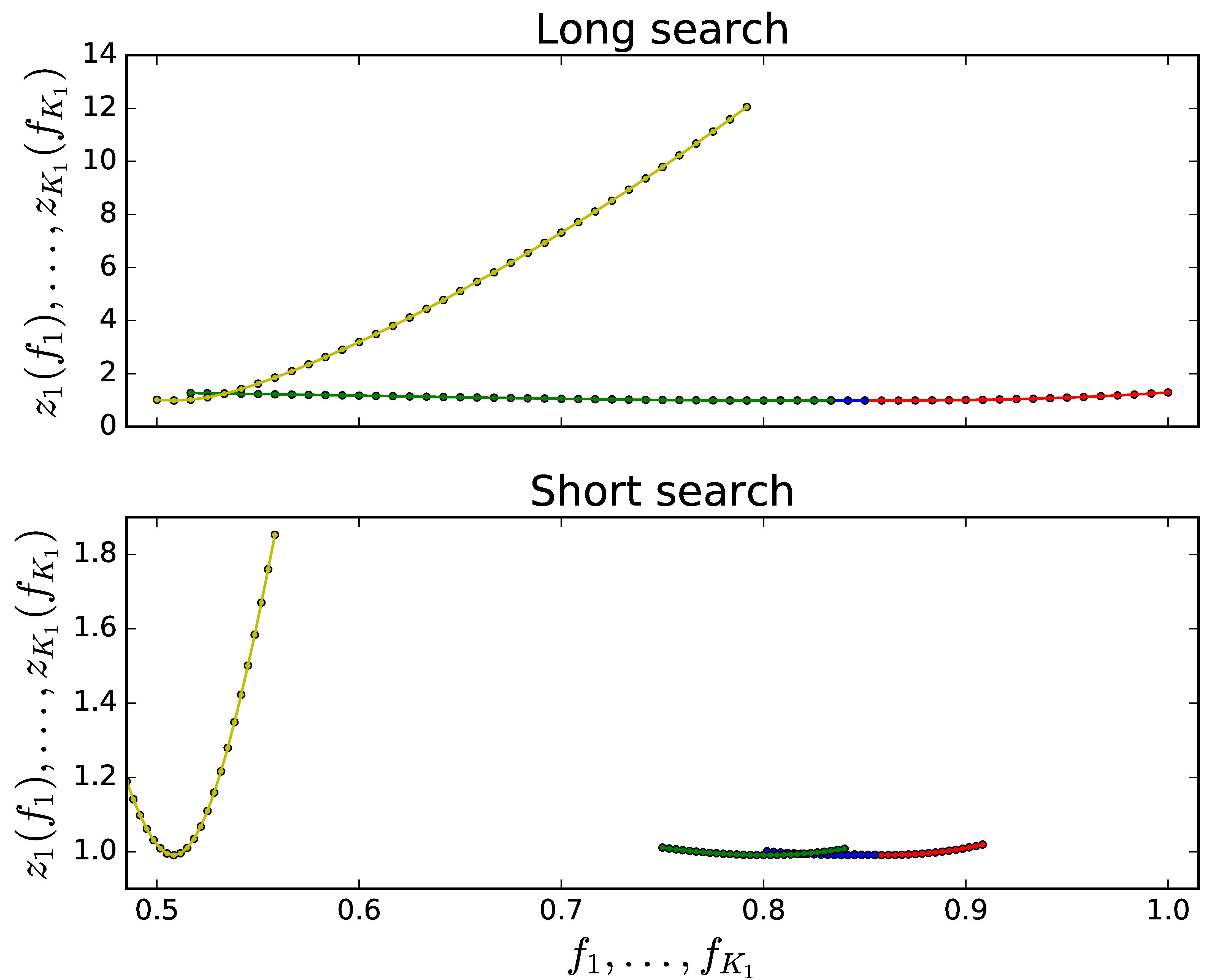}}
\end{center}
\caption{Model 27 periodograms
  $z_1(f_1), ..., z_4(f_4)$ for Table
  \ref{TableSimulatedData} data.
  Otherwise as in Fig. \ref{FigZ}.}
\label{FigTooManyOne}
\end{figure}

The periodograms in
Fig. \ref{FigTooManyOne}
are computed for
the four signal model 27
(Table \ref{TableModels}:
$K_1\!=\!4$, $K_2\!=\!1$,
$K_3\!=\!2$).
The red $z_1(f_1)$,
the blue $z_2(f_2)$
and the green $z_3(f_3)$
periodogram levels
are low and stable,
and their minima
are shallow.
Only the yellow $z_4(f_4)$
periodogram shows a
clear minimum
(Fig. \ref{FigTooManyOne}:
lower panel).
\kista{The detected periods
$P_1\!=\!1.16$,
$P_2\!=\!1.19$,
$P_3\!=\!1.25$ and
$P_4\!=\!1.97$
differ from the correct
model 19 periods
$P_1\!=\!1.10, P_2\!=\!1.40$
and $P_3\!=\!1.90$}
(Table \ref{TableSimOne}).

\begin{figure}
\begin{center}
  \resizebox{8.5cm}{!}
  {\includegraphics{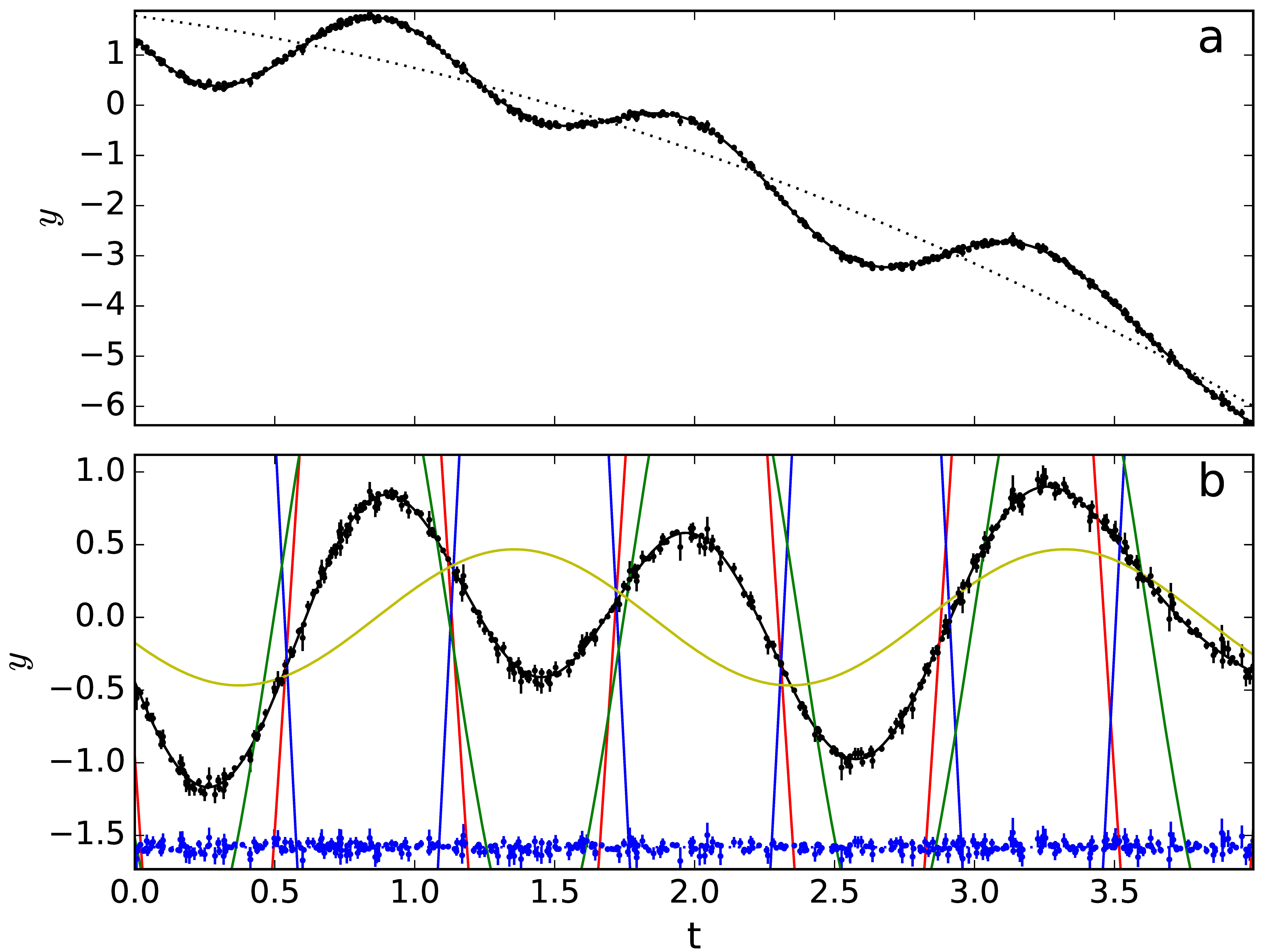}}
\end{center}
\caption{Model 27 for  Table
  \ref{TableSimulatedData} data.
  Otherwise as
  in Fig. \ref{FigModel}.}
\label{FigTooManyTwo}
\end{figure}

\begin{figure}
\begin{center}
  \resizebox{8.5cm}{!}
  {\includegraphics{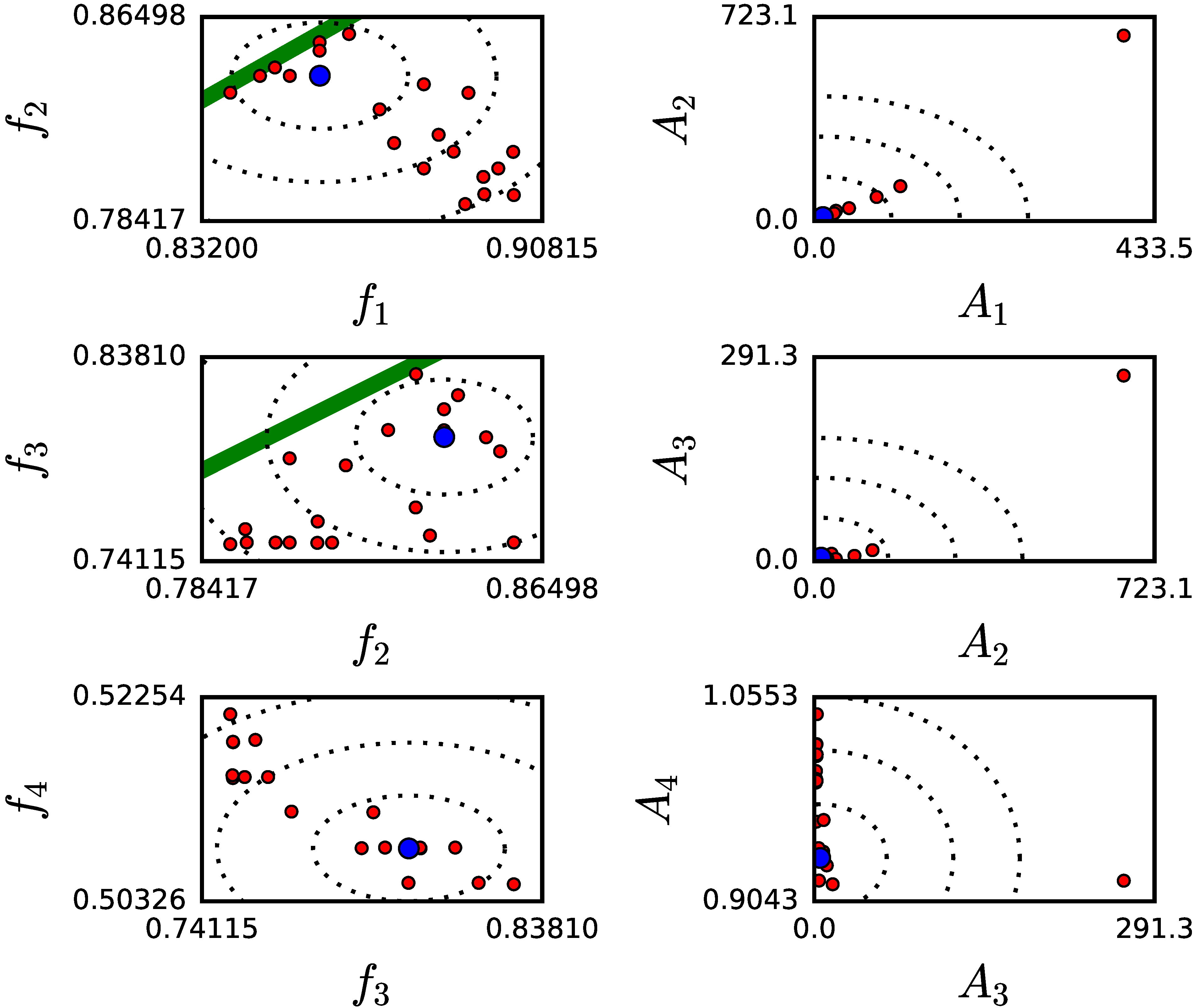}}
\end{center}
\caption{Model 27
  frequencies
  ($f_1,f_2, f_3, f_4$) and
  amplitudes ($A_1,A_2,A_3,A_4$) for
  Table
  \ref{TableSimulatedData} data.
  Otherwise as
  in Fig. \ref{FigTwoSample}.}
\label{FigTooManyThree}
\end{figure}

Model 27 ``explodes'', because
the amplitudes of
the red $h_1(t)$,
the blue $h_2(t)$ and
the green  $h_3(t)$ signals
disperse,
and only the amplitude
of the yellow $h_4(t)$
signal is
stable (Fig. \ref{FigTooManyTwo}).
We refer to this result as
\begin{itemize}
\item[] Dispersing
  amplitudes.
\end{itemize}
These dispersing
large amplitude curves
nearly cancel out
each other,
which gives a
reasonable $\chi^2\!=\!490.95$ value
(Table \ref{TableModels}). 
The bootstrap results
show that
the dotted 
frequency error lines
intersect the thick
green continuous
$f_1\!=\!f_2$
and $f_2\!=\!f_3$
diagonal lines
(Fig. \ref{FigTooManyThree}).
We refer to this as
\begin{itemize}
\item[] Intersecting frequencies.
\end{itemize}
Model 27 fails,
because the data
do not contain four signals,
but only three.
Actually,
all four signal models
consistently fail.
\kista{Nearly two thirds
  }
of the two, the three
and the four signal models
fail (Table \ref{TableModels}:
Fail=''Yes'' for 15 models out of 24).
All these failed models are just
an additional proof for that
model 19 is the best model
for Table \ref{TableSimulatedData}
simulated
data.
In fact,
we could have rejected
these failed models without ever
computing their $\chi^2$ estimates.
Furthermore,
the rejected 
one signal
models 1-8 with
very high $\chi^2$
can not
have Fail=''Yes''
or ``No''.
These one signal models
simply can not have
``intersecting
frequencies'' or
``dispersing amplitudes'',
because this
plural alternative
is impossible.
For these one signal models,
there is no need
for applying the
frequency and the amplitude
criteria,
which will be introduced later
(see Eqs. \ref{EqFreqCrit}
and \ref{EqAmpCrit}).
An unambiguous separation
between the signal 
and the trend is easiest when
the one signal model is the
correct model.

\subsection{Finding too
  few signals}
\label{SectTooFew}

\begin{figure}
\begin{center}.
\resizebox{8.5cm}{!}
{\includegraphics{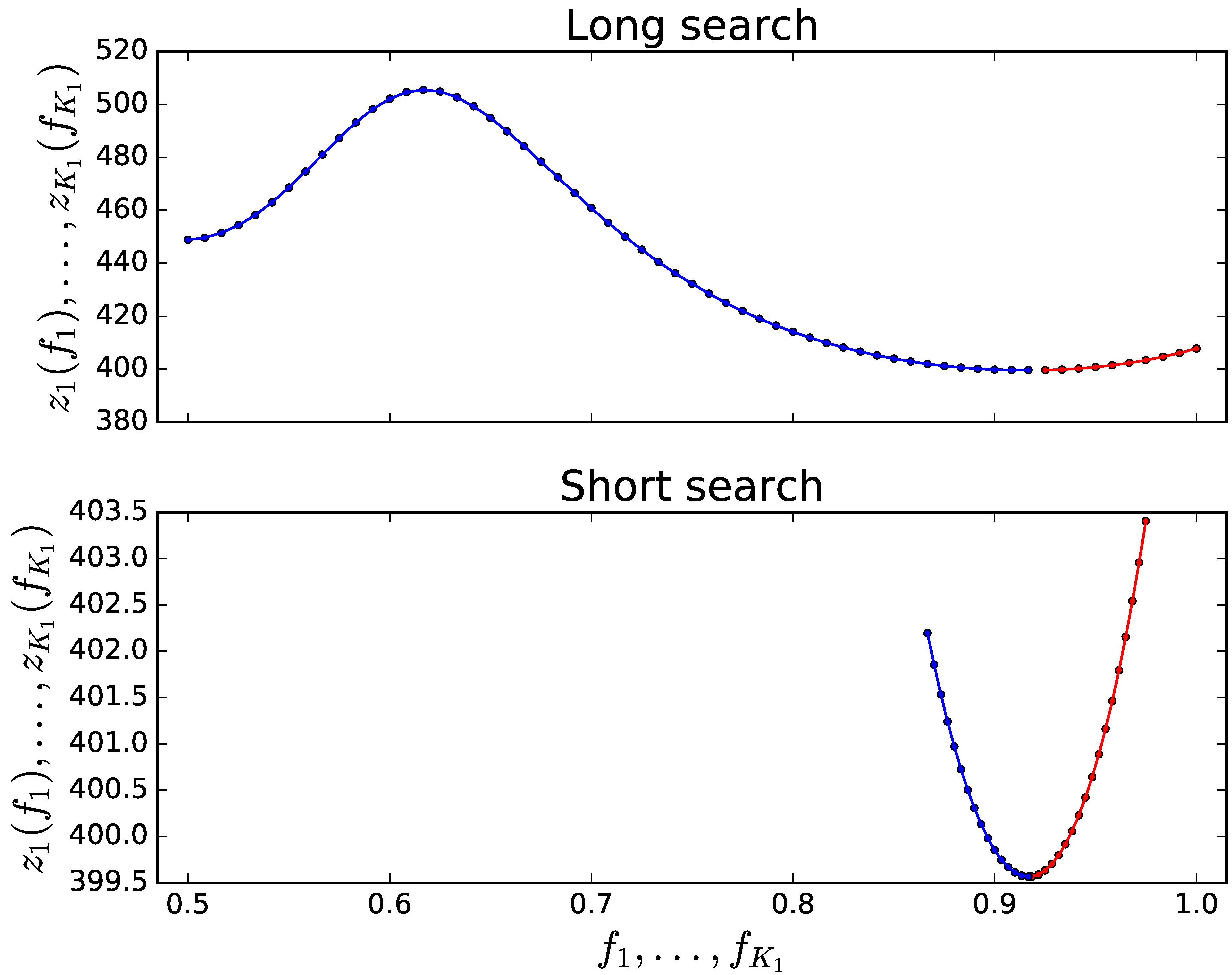}}
\end{center}
\caption{Model 9 periodograms
  $z_1(f_1)$ and $z_2(f_2)$ for Table
  \ref{TableSimulatedData} data.
  Otherwise as in Fig. \ref{FigZ}.}
\label{FigTooFewOne}
\end{figure}

\begin{figure}
\begin{center}
  \resizebox{8.5cm}{!}
  {\includegraphics{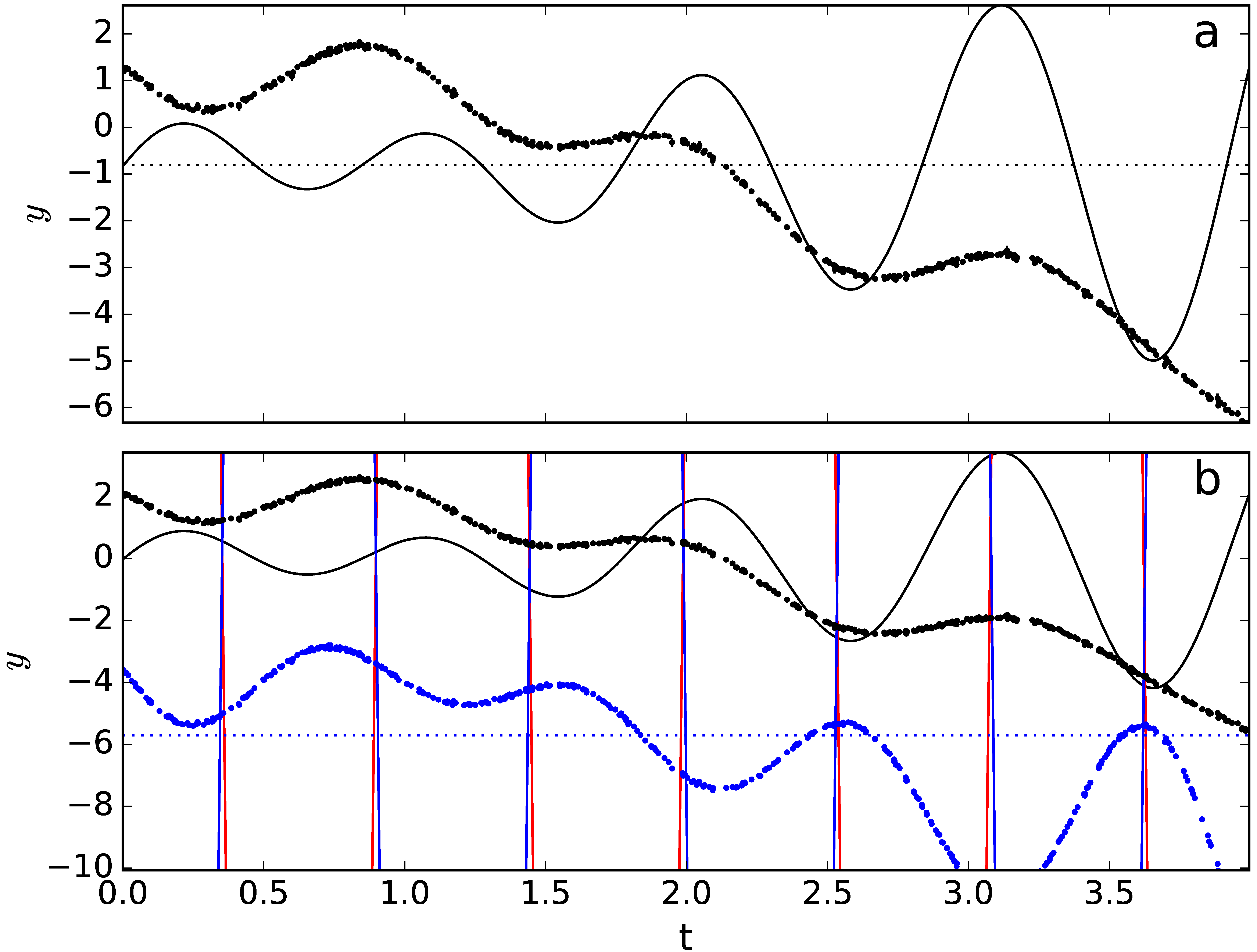}}
\end{center}
\caption{Model 9 for  Table
  \ref{TableSimulatedData} data.
  Otherwise as in
  Fig. \ref{FigModel}}
\label{FigTooFewTwo}
\end{figure}

\begin{figure}
\begin{center}
  \resizebox{8.5cm}{!}
  {\includegraphics{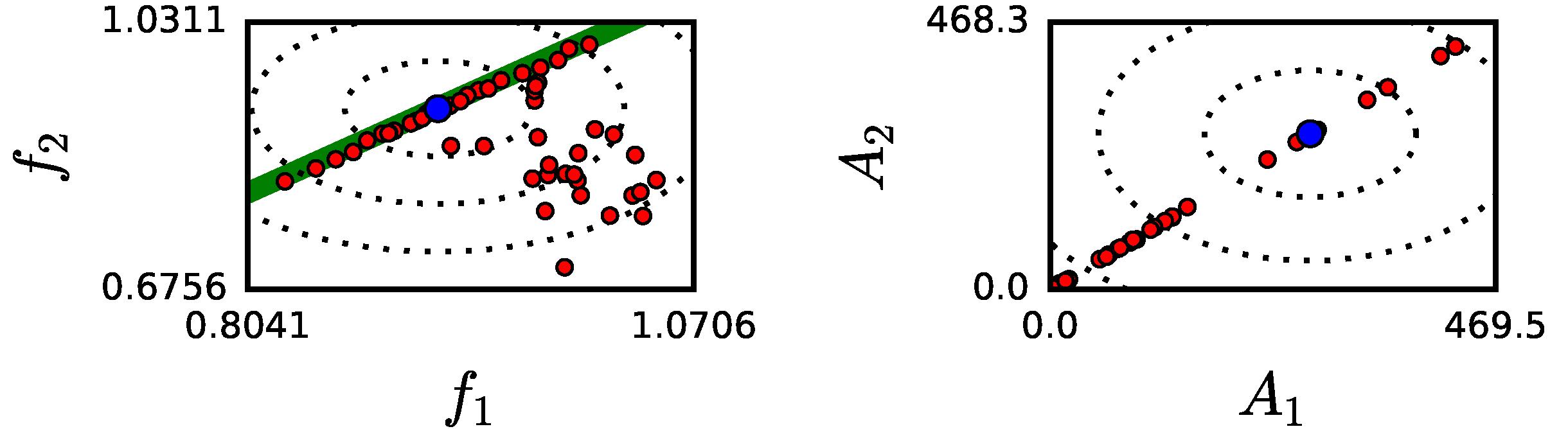}}
\end{center}
\caption{Model 9
  frequencies
  ($f_1,f_2$) and
  amplitudes ($A_1,A_2$) for
  Table
  \ref{TableSimulatedData} data.
  Otherwise as in
  Fig. \ref{FigTwoSample}.}
\label{FigTooFewThree}
\end{figure}

The simulated data of
Table \ref{TableSimulatedData}
contains three signals.
Yet, the two signal
$K_1=2, K_2=1$ and $K_3=0$ model 9
periodograms $z_1(f_1)$
and $z_2(f_2)$ 
merge, and show only the
minimum of one period
(Fig. \ref{FigTooFewOne}).
The black $g(t)$ curve
of this model 9 makes no sense
(Fig. \ref{FigTooFewTwo}a).
The red $h_1(t)$ and
the blue $h_2(t)$ signal
curves disperse, and the blue
residuals show regular
variation
(Fig. \ref{FigTooFewTwo}b).
The $f_1$ and $f_2$
frequencies intersect,
and the $A_1$ and $A_2$
amplitudes disperse
(Fig. \ref{FigTooFewThree}).
This model fails,
because the use of $K_3=0$
order $p(t)$ polynomial
totally ignores
the real trend in the data.
This idea is
supported by the fact
that all two, three and
four signal models
having $K_3=0$ consistently
fail (Table \ref{TableModels}:
models
9, 13, 17, 21, 25 and 29).

These results show that
even if two signals
are not detected in the data,
this does not mean
that the correct
number of signals
can not be three or even more.
The detection of
the correct number of signals 
depends on the selection of the
correct trend.
Wrong $p(t)$ trend can eliminate
real signals.
The results in
Table \ref{TableModels}
indicate that the false
detection of too
many signals is inprobable,
because {\it all}
four signal models
25-32 fail.
For all 32 nested models
of Table \ref{TableModels},
the \kista{false detection of too few
signals is more probable than
the false detection of
too many signals,
because the
  two signal models
  fail only two times
  out of eight
  (only models 9 and 13),
  but 
  the four signal
  models fail
  eight times out of
  eight (all models 25-32)}

\subsection{Many
  simulated models}
\label{SectManyModels}

Here, we create
artificial data
with 
{\it many} 
simulated
models
having random 
signal 
frequencies.
We show that our method
can retrieve 
the known input 
parameters of these models.
The $K_1$ simulated 
$f_i^{\star}$ frequencies
are selected from 
a uniform random distribution
\begin{eqnarray}
U(f_{\mathrm{min}},f_{\mathrm{max}},K_1)
\end{eqnarray}
between
$f_{\mathrm{min}}\!=\!1/P_{\mathrm{max}}$
and
$f_{\mathrm{max}}\!=\!1/P_{\mathrm{min}}$,
where $P_{\mathrm{min}}\!=\!1$ 
and $P_{\mathrm{max}}\!=\!2$. 
These random frequencies 
are rearranged 
into decreasing order 
$f_1^{\star} > f_2^{\star}
... >f_{K_1}^{\star}$.
They give 
$\bar{\beta}_I^{\star}=[f_1^{\star},
 f_2^{\star} ..., f_{K_1}^{\star}]$
for the simulated $g(t)$
model (Eq. \ref{Eqmodel}).

The $K_1 \times 2 K_2$
values for the
amplitudes
$B_{1,1}^{\star},$
$ C_{1,1}^{\star},$ $...,$
$ B_{K_1,K_2}^{\star}, C_{K_1,K_2}^{\star}$ 
of the simulated $h_i(t)$ signals,
as well as the $K_3+1$ values
for the coefficients
$M_0^{\star}, ..., M_{K_3}^{\star}$ 
of the simulated $p_k(t)$
polynomials,
are drawn from a uniform
random distribution
\begin{equation}
  U(-0.5,+0.5,
  K_1\times 2 K_2 + K_3+1).
\label{EqUniformBeta}
\end{equation}
The above
signal amplitudes and polynomial
coefficients give
$\bar{\beta}_{II}^{\star}$ for
the simulated $g(t)$ model
(Eq. \ref{Eqmodel}).
All free parameters 
of this simulated $g(t)$ model
are $\bar{\beta}^{\star}
=[\bar{\beta}_I^{\star},
\bar{\beta}_{II}^{\star}]$.

The simulated
$n^{\star}=500$
time points $t_i^{\star}$
are drawn from
a uniform random distribution
of Eq. \ref{EqRandomT},
where $\Delta T=4$.

The chosen signal
to noise ratio ${\mathrm{SN}}$
and the standard deviation
$s_y^{\star}$ of all 
$h(t^{\star}_i)$ give
the accuracy $\sigma_m$ of
the simulated
data (Eq. \ref{EqSigmaSim}).
The $n^{\star}$ errors
$\sigma_i^{\star}$
for the simulated
data are drawn
from the Gaussian distribution
of Eq. \ref{EqError}.

Finally,
the simulated data are
\begin{eqnarray}
  y_i^{\star}=
  g(t_i^{\star},\bar{\beta}^{\star})
  +\sigma_i^{\star}.
\label{EqSimManyData}
\end{eqnarray}

\begin{figure}
\begin{center}
  \resizebox{8.5cm}{!}
  {\includegraphics{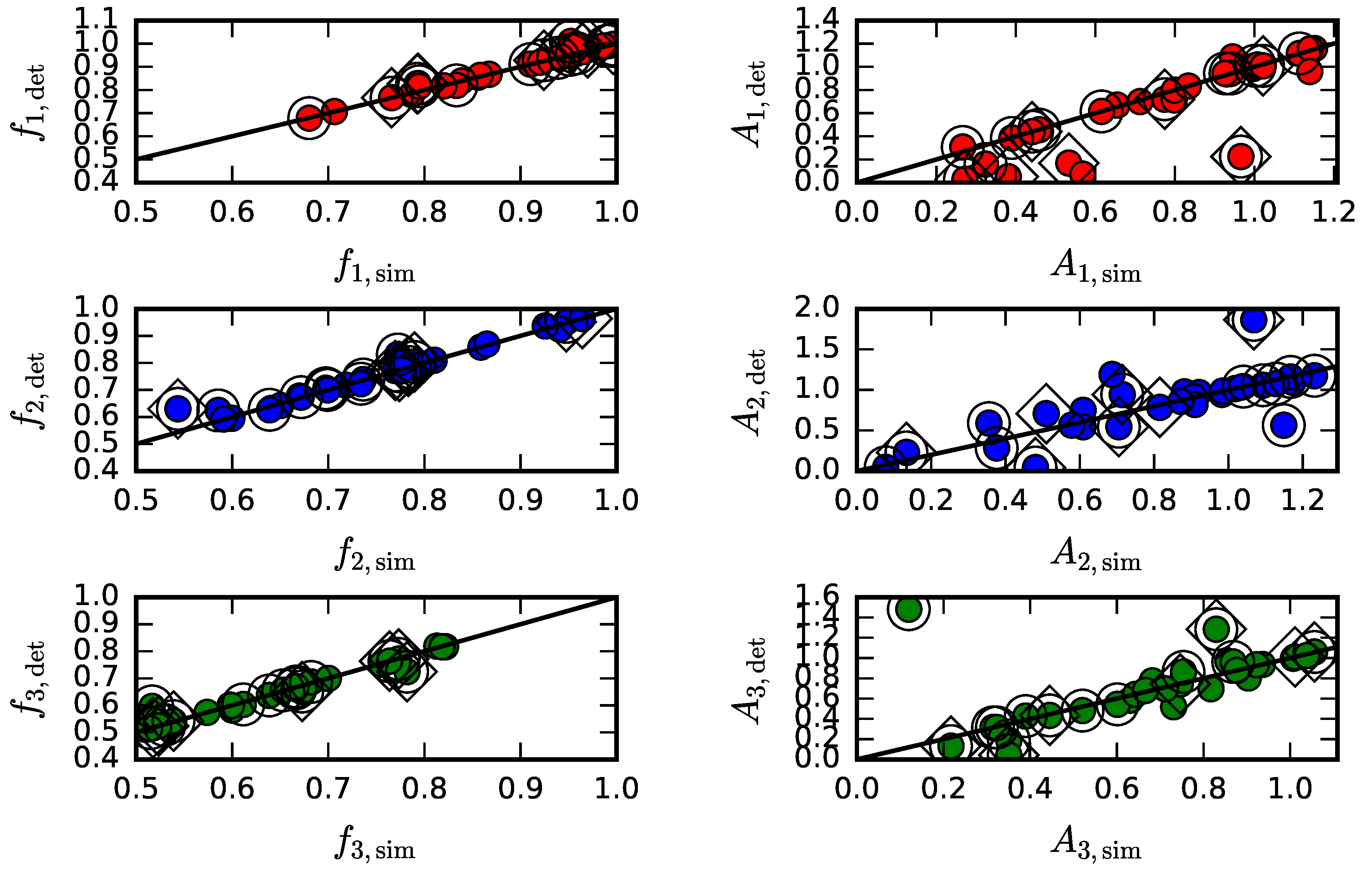}}
\end{center}
\caption{\kista{Thirty}
  simulated data
samples created
with three signal model 18.
Left hand panels show
simulated 
$f_{\mathrm{i,sim}}$ 
and detected 
$f_{\mathrm{i,det}}$ frequencies.
Right hand panels show
respective amplitudes.
Highlighted models
fulfill criteria of 
Eq. \ref{EqFreqCrit} (Transparent
diamonds)
and 
Eq. \ref{EqAmpCrit} (Transparent
circles).
Continuous lines denote
equal simulated
and detected levels.
Symbol colours are given  
in Table \ref{Tablecolour}. }
\label{FigMany}
\end{figure}

We use the three
signal
model 18 
$(K_1\!=\!3, K_2\!=\!1,
K_3\!=\!1)$ to
produce simulated
data $y_i^{\star}$
of Eq. \ref{EqSimManyData}.
Our sample size is
$n^{\star}\!=\!500$ and the
signal to noise ratio
is ${\mathrm{SN}}\!=\!100$.
The results for
\kista{thirty
model 18 simulations}
are shown in
Fig. \ref{FigMany}.
If this
DCM analysis
of ours
succeeds, 
the simulated 
frequencies $f_{\mathrm{i,sim}}$
and the detected
frequencies $f_{\mathrm{i,det}}$
in these samples
should coincide with
the continuous equal value
diagonal lines.

The transparent diamonds
in Fig. \ref{FigMany}
highlight 
models having
at least one
simulated
frequency pair
that fulfills
\begin{eqnarray}
f_{\mathrm{i,sim}}-f_{\mathrm{i+1,sim}} <
  f_{\mathrm{crit}}
  (f_{\mathrm{max}}-f_{\mathrm{min}}),
\label{EqFreqCrit}
\end{eqnarray}
  \noindent where
  $f_{\mathrm{crit}}\!=\!0.05$
  and $i=1$ or 2.
  These signal frequencies
  differ less
  than $\pm 5\%$
  in the tested frequency
  range between
  $f_{\mathrm{min}}$
  and $f_{\mathrm{max}}$.
  The models for
  these particular
  simulated
  samples
  may fail due to the
  dispersing amplitudes
  and
  the
  intersecting frequencies
  discussed in Sects. 
  \ref{SectTooMany} and
  \ref{SectTooFew}.
  As expected,
  some of these
  highlighted
  detected frequencies
  $f_{\mathrm{i,det}}$
  and amplitudes
  $A_{\mathrm{i,det}}$
  do
  deviate from the
  equal value 
  levels
  in Fig. \ref{FigMany}.

  The transparent circles
  in Fig. \ref{FigMany}
  highlight models
\begin{eqnarray}
A_i/A_{\mathrm{max}}
< A_{\mathrm{crit}},
\label{EqAmpCrit}
\end{eqnarray}
where
  $A_{\mathrm{crit}}\!=\!0.5$ and
$A_{\mathrm{max}}$ is the highest 
value of all 
signal amplitudes $A_i$
$(i=1, 2, 3)$.
For these particular
simulated samples,
signal detection becomes
more difficult,
because at least one signal
is
two times weaker than 
the strongest signal.
Again,
as expected,
the detected frequencies
$f_{\mathrm{i,det}}$
and
amplitudes
$A_{\mathrm{i,det}}$
for
some these highlighted samples
deviate from the diagonal
equal value
lines in
Fig. \ref{FigMany}.

Clearly,
the simulated
model signal 
frequencies
(Eq. \ref{EqFreqCrit}:
$f_{\mathrm{i,sim}}$)
and amplitudes
(Eq. \ref{EqAmpCrit}:
$A_{\mathrm{i,sim}}$)
determine
the success of DCM analysis.
We can confirm this important
result 
from the relative error
\begin{eqnarray}
\sigma_{\mathrm{f_i,rel}}= 
  |f_{\mathrm{i,det}}
  -f_{\mathrm{i,sim}}|/f_{\mathrm{i,sim}}.
\label{EqRelativeF}
\end{eqnarray}
It measures the
error for
the detected frequency
$f_{\mathrm{i,det}}$ in
the units of the simulated
frequency $f_{\mathrm{i,sim}}$.
We compute the mean of
relative errors
$\sigma_{\mathrm{f_i,rel}}$
for $m=100$ simulated
samples $y_i^{\star}$
produced with
model 18,
$n^{\star}=500$ and
$\mathrm{SN}=100$
(Eq. \ref{EqSimManyData}).
The results 
are given in
Table \ref{TableFreqError}
(Lines 1-3).
The mean of
relative error $\sigma_{\mathrm{f_i,rel}}$
decreases when models
fulfilling criterion of
Eq. \ref{EqFreqCrit} are
\kista{removed $(m=100 \rightarrow 69)$.}
It 
decreases
even more
when models
fulfilling criterion of
Eq. \ref{EqAmpCrit}
are \kista{also removed
  $(m=69 \rightarrow 48)$.}
These general results
are consistently confirmed
with
doubled signal to noise ratio
$\mathrm{SN}=100 \rightarrow200$
(Table  \ref{TableFreqError}:
Lines 4-6)
and
doubled sample size
$n^{\star}=500 \rightarrow 1000$
(Table  \ref{TableFreqError}:
Lines 7-9).
This confirms that
our DCM can
detect the correct
frequencies when

\begin{figure}
\begin{center}
  \resizebox{6.5cm}{!}
  {\includegraphics{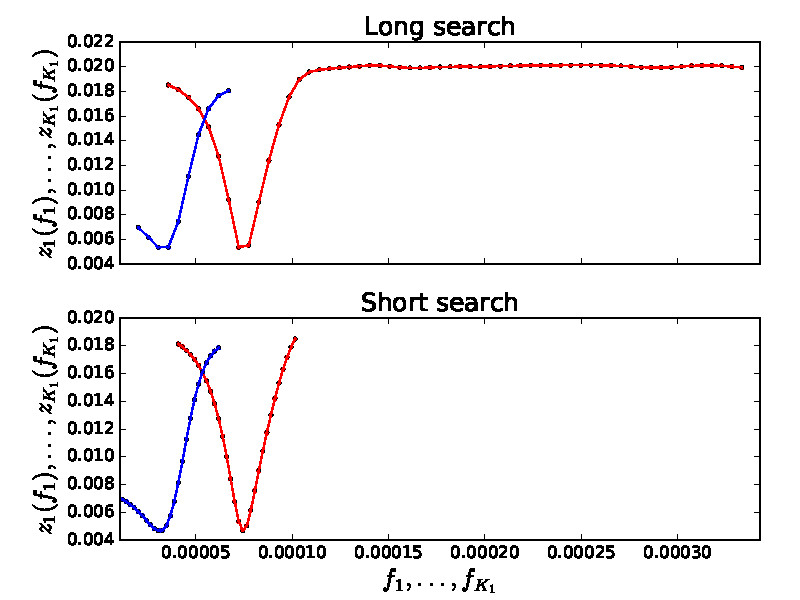}}
\end{center}
\caption{Periodograms
  $z_1(f_1)$ and $z_2(f_2)$ for XZ And 
  data.
  Otherwise as in Fig. \ref{FigZ}.}
\label{FigRealCase1}
\end{figure}

\begin{figure}
\begin{center}
  \resizebox{6.5cm}{!}
  {\includegraphics{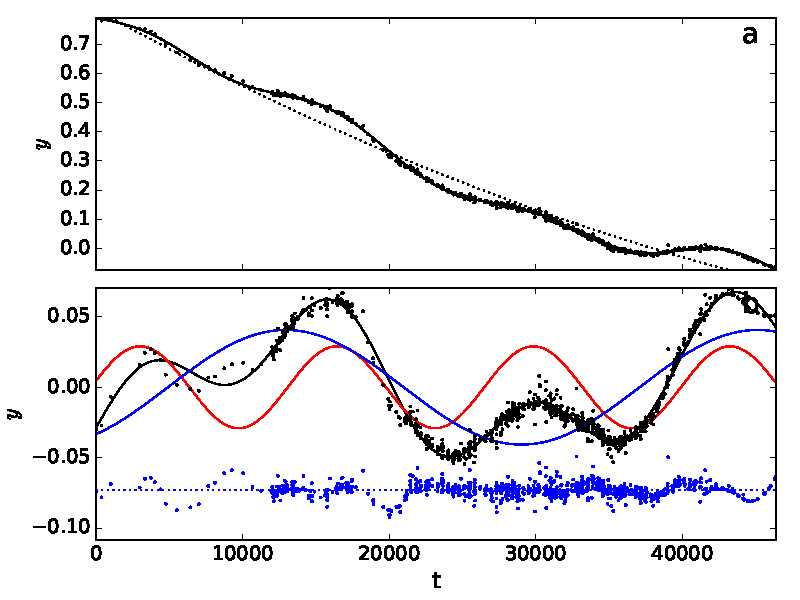}}
\end{center}
\caption{Model for XZ And data.
  Otherwise as
  in Fig. \ref{FigModel}.}
\label{FigRealCase2}
\end{figure}

\begin{table}
\caption{Mean of relative errors 
  $\sigma_{\mathrm{f_1,rel}}$,
  $\sigma_{\mathrm{f_2,rel}}$ 
  and
  $\sigma_{\mathrm{f_3,rel}}$ 
  for one
hundred model 18 simulations
with
$n^{\star}=500$ and SN=100.
Lines 1-3 give results
for all models $(m=100)$,
when
Eq. \ref{EqFreqCrit}
criterion models are
\kista{removed  $(m=100
  \rightarrow 69)$},
and 
when
Eq. \ref{EqAmpCrit} criterion
models are also
\kista{removed $(m=69\rightarrow 48)$}.
Signal to noise ratio
${\mathrm{SN}}$ 
is doubled on next three lines.
Sample size $n^{\star}$
is doubled on
next three lines.} 
\begin{center}
\begin{tabular}{clcccr}
\hline
  &                                   & \multicolumn{4}{c}{$n^{\star}=500, {\mathrm{SN}=100}$}                  \\
Line &   Samples                      & $\sigma_{\mathrm{f_1,rel}}$  &   $\sigma_{\mathrm{f_2,rel}}$  &   $\sigma_{\mathrm{f_3,rel}}$  & $m$\\
  \cline{3-6}
1 & All                                       &  0.012                  &  0.029                  & 0.0090                    & 100 \\
2 & Eq. \ref{EqFreqCrit}                      &  0.0085                 &  0.013                  & 0.0065                    &  69 \\
3 & Eqs. \ref{EqFreqCrit} and \ref{EqAmpCrit} &  0.0030                 &  0.011                  & 0.0051                    &  48 \\
\hline
  &                                        & \multicolumn{4}{c}{SN doubled: $n^{\star}=500, {\mathrm{SN}=200}$ }                       \\
Line &  Samples                            & $\sigma_{\mathrm{f_1,rel}}$  &   $\sigma_{\mathrm{f_2,rel}}$  &   $\sigma_{\mathrm{f_3,rel}}$  & $m$\\
  \cline{3-6}

4 &  All                                      &   0.0039                 &  0.014                    &   0.011                    & 100 \\
5 &  Eq. \ref{EqFreqCrit}                     &   0.0036                 &  0.0083                   &   0.0082                   &  76 \\
6 & Eqs. \ref{EqFreqCrit} and \ref{EqAmpCrit} &   0.0019                 &  0.0036                   &   0.0032                   &  37 \\
\hline
   &                                      & \multicolumn{4}{c}{$n^{\star}$ doubled: $n^{\star}=1000, {\mathrm{SN}=100}$ }                   \\
Line & Samples                          & $\sigma_{\mathrm{f_1,rel}}$  &   $\sigma_{\mathrm{f_2,rel}}$  &   $\sigma_{\mathrm{f_3,rel}}$  & $m$\\
  \cline{3-6}
7 &  All                                      &  0.010                   & 0.019                     &  0.0050                    & 100 \\
8 & Eq. \ref{EqFreqCrit}                      &  0.0064                  & 0.015                     &  0.0049                   & 77 \\
9 & Eqs. \ref{EqFreqCrit} and \ref{EqAmpCrit} &  0.0034                  & 0.0077                    &  0.0041                    & 40 \\
\hline
\end{tabular}
\end{center}
\label{TableFreqError}
\end{table}

\begin{enumerate}

\item Signal frequencies
  are not too close
(Eq. \ref{EqFreqCrit}).

\item None of the signal
  amplitudes is too
weak (Eq. \ref{EqAmpCrit}).

\item Sample size $n$ and signal
  to noise ratio
  ${\mathrm{SN}}$
  are sufficient
  (Table \ref{TableFreqError}).

\end{enumerate}

\noindent
If these correct frequencies
are detected,
the values
for the remaining other model 
parameters
will also be correct,
because
linear modelling
is always unambiguous.
Nevertheless,
failing to detect 
even a single correct
frequency can seriously 
mislead the period analysis
(e.g. Figs. \ref{FigTooManyTwo}
and \ref{FigTooFewTwo}).

\section{Real use case}
\label{SectXZAnd}

We also
  provide one example of
  preliminary
  real case use of DCM.
  
Periodic changes occur
in the observed (O) minus the computed (C)
eclipse epochs of binaries.
The most probable causes for
such periodicities are
a third body \citep[e.g.][]{Li18},
a magnetic activity cycle \citep[e.g.][]{App92}
or
an apsidal motion \citep[e.g.][]{Bor05}.
Recently, \citet{Haj19} searched for 
third bodies in a large sample of about
80~000 eclipsing binaries. 
The model in their one-dimensional
period analysis of O-C data
was a pure sinusoidal signal
superimposed on a second order
polynomial \citep[][Eq. 11]{Haj19}.
They discovered 992
hierarchial triple systems, but only
four candidates possibly having a fourth body.
In other words, the probability for
finding a fourth body in their large
sample was about $4/80~000=0.00005$.

Our data are
the observed (O) minus the computed (C)
primary minimum epochs of XZ And,
which were retrieved from
Lichten\-knec\-ker-Database of the
BAV\footnote{XZ And data were retrieved
  from \Link{https://www.bav-astro.eu/index.php/veroeffentlichungen/lichtenknecker-database/lkdb-b-r}
{Lichten\-knec\-ker-Database of the BAV}}
in September 2019.
The primary (A4~IV-V, $3.2 M_{\odot}$, $2.4 R_{\odot}$)
and
the secondary (G~IV, $1.3 M_{\odot}$, $2.6 R_{\odot}$)
of this binary orbit each other
during $P_{\mathrm{orb}}=1.357$ days
\citep{Dem95}.
Our O-C data values had been computed from the
ephemeris 
\begin{eqnarray}
\mathrm{JD~}2452500.5129 + 1.35730911{\mathrm{E}}
\label{EqEphe}
\end{eqnarray}
\citep{Kre04}.
Since there already was evidence for the possible
presence of a third body in \target
~\citep[][]{Dem95,Man16,Cha19},
we applied DCM to the O-C data of \target.
The DCM periodograms for the
$g(t,2,1,2)$ model are shown
in Fig. \ref{FigRealCase1}.
We detect the periods
$P_1=13418^{\mathrm{d}}=$$37^{\mathrm{y}}$ and
$P_2=32192^{\mathrm{d}}=$$88^{\mathrm{y}}$.
These two signals are shown
in Fig. \ref{FigRealCase2}.
For example,
\citet{Dem95} and \citet{Cha19}
have also detected 
the $P_1$ periodicity.
It seems that our DCM can easily
detect evidence for the possible
presence of a fourth body in \target.
This result
illustrates the potential of DCM in
detecting many signals superimposed
on unknown polynomial trends.

From the DCM point of view,
the most important currently
unsolved questions are

\begin{enumerate}

\item How many $K_1$ signals do
  these data contain?
\item What is the correct $K_3$ trend?
\end{enumerate}

\noindent
Since the aim of the current paper
is to present DCM, the final
results of this preliminary real
use case will be presented in a
future work.
For this reason, no DCM files of this
case are published here.

\section{Discussion}

The main point of
  DCM is that 
  our
  periodic 
  non-linear
  model
  becomes linear when
  the grid of 
  constant tested frequencies
  is fixed.
  All analysis
  results 
  become unambiguous,
  which guarantees
  the success of DCM.
  Actually,
  we can now present a general numerical
  solution for {\it any} non-linear
  $g(t,\bar{\beta})$ model.
  This non-linear model
   may be
  periodic,
  aperiodic,
  or 
  combination of
  both, like the DCM model.
  Our simple recipe is
  \begin{enumerate}
  \item Divide the free parameters $\bar{\beta}$
  to two parts:
  \subitem a: Those that make the model nonlinear = $\bar{\beta}_I$
  \subitem b: The rest of the free parameters = $\bar{\beta}_{II}$
  \item Fix the tested $\bar{\beta}_I$ grid.
  \item Test all reasonable linear models.
  \item Identify the best model among these models.
  \item Solve the model parameter errors
        with
        bootstrap.

  \end{enumerate}
   
  Another main point
  of DCM is
  the $z$ test statistic
  symmetry
  in the $K_1$-dimensional 
  frequency space.
Without this symmetry,
our period search would
literally resemble
the search for a needle
in a haystack for
higher number of signals.
For example,
the six
signal models have
$K_1!=6! = 720$ symmetries,
which give the same
number equally good alternative
$z$ periodogram minima in
six-dimensional 
frequency space.  
This $z$ symmetry allows us
  to test only a single
  frequency combination,
  and to get 
  rid of the other
  irrelevant 
  $K_1!-1$ frequency
  combinations. 
  Whatever the correct real 
  frequency values may be,
  they can always be 
  rearranged into
  a decreasing
  order $f_1 > f_2 > ... >f_{K_1}$.
  Therefore, we test only
  the combinations of all
  those one-dimensional
  frequency intervals that 
  do not overlap.
  We never have to bother 
  about the rest
  $(K_1!-1)/K_1!$
  of the entire frequency space,
  because nothing new can be 
  found out there.
 
  All periodogram minima
  of model 19 are steep
  in Fig. \ref{FigZ},
  which means that
  $\chi^2$
  with
  the initial
  estimate 
  $\bar{\beta}_{\mathrm{Initial}}$
  is already
  very close
  to its possible minimum
  value
  before
  the non-linear
  minimization iteration
  of Eq. \ref{EqInitial}
  even begins.
  The $z_i(f_i)$ periodograms
  in all
  Figs. \ref{FigZ},
  \ref{FigTooManyOne}
  and \ref{FigTooFewOne}
  display
  no sudden jumps, because
  there is strong correlation
  between the $\chi^2$
  values for tested
  frequencies close to each other.
  If the grid
  of tested constant frequencies
  is already
  sufficiently dense,
  there is no sensible
  ``escape'' away 
  from
  the minima of these continuous,
  stable and unambiguous 
  $z_i(f_i)$ periodogram
  curves for linear models.
  Thus,
  the non-linear
  iteration of Eq. \ref{EqInitial}
  can not very much
  improve the $\chi^2$ estimate,
  because the search
  for the best
  model solution is
  already
    nearly over.
  For example, the  $\chi^2$
  estimates for all 32
  models of Table \ref{TableModels}
  are practically
  the same with,
  or without, the
  non-linear iteration
  alternative\footnote{\MyDat
   ~alternatives
  \PR{NonLinear = 1} =Yes and
  \PR{$\neq$ 1} = No.}
   when the
  frequency grid parameters are
  fixed to $n_{\mathrm{L}}=60$,
  $n_{\mathrm{S}}= 30$ and $a=0.20$.
  Accurate
  model parameter estimates,
  including the frequencies,
  can already
  be obtained with linear models
  when the tested frequency grid is
  not too sparse. 
  We conclude that
  the nonlinear iteration 
  of Eq. \ref{EqInitial}
  is not always needed.  
  However, the word ``Discrete'' 
  in our DCM abbreviation
  could be replaced
   with the word  
   ``Continuous'' when 
   this non-linear
   iteration of
   Eq. \ref{EqInitial} 
   is applied.

There are 
  correlations between
  the signal frequencies 
  and amplitudes
  of the correct model 19
  (Fig. \ref{FigTwoSample}).
  If an estimate for 
  even one of these
  parameters shifts 
  away from the correct
  value, the remaining 
  other estimates tend
  to compensate this shift 
  with their own shifts
  away from their 
  correct values.
  These shifts may mislead
  the DCM period analysis,
  or at least
  increase the bootstrap error
  estimates,
  if the tested frequency 
  grid is too
  sparse in the long or
  the short search, 
  or in the bootstrap. 
  This possibly
  misleading
  effect can be 
  eliminated with
  denser tested frequency
  grids,
  but then
  the
  detection of many signals
  requires a lot of computation time,
  because
  the total number of tested 
  frequency combinations
  is proportional to 
  $n_{\mathrm{L}}^{K_1}$
  and $n_{\mathrm{S}}^{K_1}$.
  However, 
  this ``wasted''\footnote{For example,
  an ordinary PC computes the four signal model 27
  and its thirty bootstrap
  rounds in about three days
  (Figs. \ref{FigTooManyOne}-\ref{FigTooManyThree}).
  It takes about the same time
  to compute all results in our
  Table \ref{TableFreqError}.}
  computation time
  becomes irrelevant,
  if the correct 
  frequencies are detected,
  because the results for all
  other model parameters become
  unambiguous.
  The patience required 
  in testing all possible
  parameter combinations,
  as well as all
  reasonable linear models,
  is amply rewarded. 

    We identify the best model for
    the data
    among all 32 alternative nested 
    models with
    the simple $\chi^2$-test
    and/or the standard Fisher-test
    (Sect. \ref{SectIdentify}).
    The correct model 19 has
    $p=12$ free parameters.
    We can establish this required
    minimum
    number of free parameters with
    absolute certainty,
    because 
    the critical levels for
    all fifteen models having less
    than $p=12$ free
    parameters are below
    the computational accuracy
    (Table \ref{TableModels}:
    $Q_{F}<10^{-16}$).
    The $\chi^2$ and/or $Q_F$ values
    for all sixteen remaining
    alternative    
    $p\ge12$ models
    confirm 
    that model 19
    is the best
    model.
    Furthermore,
    model 19 is certainly better
    than 
    the 
    \kista{fifteen} failed
    many signal DCM
    models,
    which can be easily
    identified
    from their dispersing
    amplitudes and intersecting
    frequencies 
    (Sects. \ref{SectTooMany}
     and \ref{SectTooFew}).

     While DFT can detect
       only pure sinusoidal signals,
       our DCM can also detect more complicated
       $h_i(t)$ signals.
       Except for
       the trivial $n+1 \ge p$ condition,
       there is no theoretical
       $K_2$ upper limit
       for the signal order that can be
       detected
       with the DCM.
       Our \PR{dcm.py} code
       can detect only 
       $K_2=1$ and $K_2=2$ order signals.
       However, the users of our  code
       should be aware of the
       problems arising
       in the DCM period search
       for higher order
       $K_2>1$ signals.
       These signals are sums of 
\begin{eqnarray}
  h_{i,j}(t)=B_{i,j}  \cos{(2\pi j f_i t)}
  +C_{i,j} \sin{(2 \pi j f_i t)},
\end{eqnarray}
\noindent
where $j=1, ..., K_2$ and $P_i=1/f_i$.
We denote the peak to peak amplitudes
of these $h_{i,j}(t)$ signals with $A_{i,j}$.
The following problems arise with these
higher order models:

\begin{itemize}

\item[1.] Let us assume that the
  second order $K_2=2$
  model is the correct model
  for the data.

  \item[1a.]
    If the 
    $A_{i,1}/A_{i,2}$ ratio of the
    $h_i(t)$ signal
    approaches zero,
  the $K_2\!=\!2$ order model
  may detect the correct 
  $P_i$ period, or the wrong half $P_i/2$ period.
  Both of these periods are equally good
  for $A_{i,1}/A_{i,2}=0$.

  \item[1b.]
    If the $h_i(t)$ signal ratio $A_{i,2}/A_{i,1}$
    approaches zero,
  the $K_2=2$ order model
  may detect the correct period
  $P_i$, or the wrong double period $2P_i$.
  For $A_{i,2}/A_{i,1}=0$, 
  both periods are equally good.

  \item[1c.]
  If a wrong model,
  like the $K_2=1$ model,
  is applied to these data, 
  the correct $P_i$ period may, or may not,
  be detected.
  Then again, the
  one-dimensional
  DTF analysis based on
  this $K_2=1$ model also
  suffers from these half and
  double period 1a-c problems
    \citep[e.g.][]{Rei13A}.

  \item[2.] Evidently, more complicated problems
  than the above 1a-c problems arise
  with the higher $K_2>2$ order signals.
  We emphasize that an unambiguous
  solution
  for the these problems {\it directly} 
  from the periodograms of a {\it single}
  model can fail,
  because this model
  may not be the correct model
  for the data.
  Nevertheless,
  there is an unambiguous solution
  for these problems {\it afterwards}:
  the Fisher-test comparison
  of {\it many} alternative
  nested models.
  It would have been possible
  for us to code
  a DCM version that first tests all
  chosen $K_1$, $K_2$ and $K_3$
  model
  combinations, and then performs
  the above Fisher-tests,
  like the comparison of 32 nested
  models in our Table \ref{TableModels}.
  We decided 
  not to code this
  tedious alternative into
  our current
  \PR{dcm.py} version, because
  the users can compare
  the DCM results for different nested
  models
  with our \PR{fisher.py} program.

\item[3.] For all these higher
    $K_2>2$ order
    complex signals,
    the probability
    for detecting a wrong period $P_i$
    for any $h_i(t)$ signal
    also depends on
    the quality $(\sigma_i)$
    and
    the quantity $(n)$
    of data,
    as well as on the frequencies of the
    other $K_1-1$ signals (Eq. \ref{EqFreqCrit}),
    and
    the
    amplitudes of these other
    signals (Eq. \ref{EqAmpCrit}).

\item[4.] There is also one computational aspect,
  why we decided not to code the
  cases $K_2>2$ into \PR{dcm.py}.
  It would be easy to compute 
  the numerical bootstrap error
  estimates for the signal
  periods and amplitudes
  of higher order $K_2>2$ models.
  However, the bootstrap
  error estimates
  for the minimum and the maximum
  epochs would not be easy to compute
  for these $K_2>2$ models.
  For example, a secondary minimum
  may be present or absent in some
  $K_2=2$ model bootstrap samples.
  Or, the primary 
  and the secondary minima may switch in
  some bootstrap samples,
  if the
  double wave of the $K_2=2$ model
  has two equally deep
  minima.
  In fact, we had already solved these
  $K_2=2$ model primary and secondary
  minimum epoch
  problems\footnote{\PR{HarmonicEpochs}
    subroutine of \PR{dcm.py}
    sorts the bootstrap
    primary
    and secondary minimum epochs.}
  earlier
  \citep[e.g.][Figs. 2 and 4]{Jet99}.
  The bootstrap solutions for these minimum
  and maximum epoch errors are
  much more complicated for $K_2>2$ models.
  
\end{itemize}

 DCM solves these tasks more
    {\it directly} than DFT:

\begin{enumerate}

    \item One signal
      data without trends: \\
      DFT finds the correct period.
      Then the data
      are modelled with a
      sinusoid having this
      period.
      DCM  achieves this
      {\it directly} with
      the $g(t,1,1,0)$
      model.

    \item One signal
     data with trends: \\
    After trend removal,
    DFT may, or may not,
    find the
    correct period. Then
    a sinusoid with this
    period
    is fitted
    to the detrended data.
    DCM  achieves
    this {\it directly} 
    with the $g(t,1,1,K_3)$
    model for any $K_3$:th order
    polynomial trend.
    
  \item Many signal data
    with trends: \\
    After trend removal, 
    the DFT
    pre-whitening technique
    may, or may not,
    determine the
    correct sequence
    of periods one
    after another.
    Then sinusoids having
    these periods are fitted
    to the detrended data.
    DCM achieves
    this {\it directly}
    for
    any order of
    polynomial trends,
    any number of signals,
    and
    any harmonic order signals,
    including one harmonic order
    pure sinusoidal signals.
    
\end{enumerate}

\noindent
In all these cases 1--3,
    DFT and DCM both
    obtain the final result
    by minimizing the $\chi^2$
    test statistic.
    DCM can find the global
    $\chi^2$ minimum,
    if {\it all} differences
    between signal frequencies are
    not too small
    (Eq. \ref{EqFreqCrit}),
    and {\it none} of the signal 
    amplitudes is too low
    (Eq. \ref{EqAmpCrit}). 
    We show that
    in this case the
    correct
    simulated
    frequencies can be
    detected,
    and their accuracy
    is only improved
    for higher signal to
    noise ratio ${\mathrm{SN}}$
    and larger sample size $n$
    (Fig. \ref{FigMany},
    Table \ref{TableFreqError}).
    When these correct
    frequencies are detected,
    all other model parameters
    are also
    correct,
    because their linear 
    least squares fit
    solutions are unambiguous.

 \section{Conclusions}

 The frequently
 applied Discrete
  Fourier Transform (DFT)
  can detect
  periodicity
  in unevenly spaced
  data.
  Unambiguous
  signal
  detection
  succeeds
  only
  if the data contains no
  trends and
  a sinusoid is
  the correct model
  \citep{Lom76,Sca82,Zec09}.
  DFT
  can not 
  {\it directly} detect
  many signals
  superimposed on unknown
  trends,
  but our
  Discrete
  Chi-Square Method (DCM) can.
  Our model for the data
  is the sum $g(t)=h(t)+p(t)$,
  where $h(t)$ contains
  the signals
  and $p(t)$ is the
  polynomial trend.
  The former periodic part 
  repeats itself,
  but the latter aperiodic part
  does not.
  Our $g(t)$ model
  is {\it non-linear},
  but it becomes {\it linear}
  when the frequencies of $h(t)$
  are fixed to their constant
  numerical tested
  frequency grid values.
  These linear models give
  {\it unambiguous} results.
  We spoil the fun of
  traditional
  time series analysis with
  our brute numerical approach,
  because we test all possible
  free parameter values
  for all reasonable
  linear models.
  We can also identify
  the best model for the
  data among all alternative
  nested
  models, and show {\it when} the
  correct frequencies can be detected
  (Eqs. \ref{EqFreqCrit}
  and \ref{EqAmpCrit}).
  If these detected
  frequencies are correct,
  all other model
  parameters are also correct.

  Anyone can 
  test
  our DCM code, but
  just like any
  any other
  period finding method code,
  it has its
  statistical limitations.
  Since there will always be
  challenging
  problems with real data,
  we also code
  the DCM alternative for
  analysing simulated
  data\footnote{The \MyPro ~program
    can simulate data similar to the
    users own real data
    with two
    \MyDat ~alternatives:
    \PR{RealData $\neq$ 1} and \PR{SimMany $\neq 1$} (Mode 2),
     or 
    \PR{SimMany = 1} (Mode 3).
    We give detailed instructions
    about this possibility
    in our Appendix.}
  similar
  to the users' own real data.

  We have now formulated,
  tested and coded DCM.
  However, we leave the
  tedious comparison between
  DCM and DFT to our next study.


\acknowledgements
We thank Dr. Karri Muinonen for
his comments
about numerical nonlinear least
squares iteration routines.
We also thank Dr. Thomas Hackman
for his comments.

\bibliographystyle{apj}


\begin{thebibliography}{25}
\expandafter\ifx\csname natexlab\endcsname\relax\def\natexlab#1{#1}\fi

\bibitem[{{Aigrain} {et~al.}(2017){Aigrain}, {Parviainen}, {Roberts}, {Reece},
  \& {Evans}}]{Aig17}
{Aigrain}, S., {Parviainen}, H., {Roberts}, S., {Reece}, S., \& {Evans}, T.
  2017, \mnras, 471, 759

\bibitem[{{Andrae} {et~al.}(2010){Andrae}, {Schulze-Hartung}, \&
  {Melchior}}]{And10}
{Andrae}, R., {Schulze-Hartung}, T., \& {Melchior}, P. 2010, arXiv e-prints,
  arXiv:1012.3754

\bibitem[{{Applegate}(1992)}]{App92}
{Applegate}, J.~H. 1992, \apj, 385, 621

\bibitem[{Barlow(1993)}]{Bar93}
Barlow, R. 1993, Statistics: A Guide to the Use of Statistical Methods in the
  Physical Sciences, Manchester Physics Series (Wiley)

\bibitem[{{Borkovits} {et~al.}(2005){Borkovits}, {Forg{\'a}cs-Dajka}, \&
  {Reg{\'a}ly}}]{Bor05}
{Borkovits}, T., {Forg{\'a}cs-Dajka}, E., \& {Reg{\'a}ly}, Z. 2005,
  Astronomical Society of the Pacific Conference Series, Vol. 333, {\rm The
  combined effect of the perturbations of a third star and the tidally forced
  apsidal motion on the O--C curve of eccentric binaries}, ed. A.~{Claret},
  A.~{Gim{\'e}nez}, \& J.~P. {Zahn}, 128

\bibitem[{{Chap\-lin}(2019)}]{Cha19}
{Chap\-lin}, G.~B. 2019, Journal of the American Association of Variable Star
  Observers (JAAVSO), 47, 222

\bibitem[{{Demircan} {et~al.}(1995){Demircan}, {Akalin}, {Selam}, {Derman}, \&
  {Mueyesseroglu}}]{Dem95}
{Demircan}, O., {Akalin}, A., {Selam}, S., {Derman}, E., \& {Mueyesseroglu}, Z.
  1995, \aaps, 114, 167

\bibitem[{Draper \& Smith(1998)}]{Dra98}
Draper, N.~R., \& Smith, H. 1998, Applied Regression Analysis (John Wiley {\&}
  Sons, Inc.)

\bibitem[{{Efron} \& {Tibshirani}(1986)}]{Efr86}
{Efron}, B., \& {Tibshirani}, R. 1986, Statistical Science, 1, 54

\bibitem[{{Hajdu} {et~al.}(2019){Hajdu}, {Borkovits}, {Forg{\'a}cs-Dajka},
  {Sztakovics}, {Marschalk{\'o}}, \& {Kutrov{\'a}tz}}]{Haj19}
{Hajdu}, T., {Borkovits}, T., {Forg{\'a}cs-Dajka}, E., {et~al.} 2019, \mnras,
  485, 2562

\bibitem[{{Jetsu} \& {Pelt}(1999)}]{Jet99}
{Jetsu}, L., \& {Pelt}, J. 1999, \aaps, 139, 629

\bibitem[{{Jetsu} {et~al.}(2013){Jetsu}, {Porceddu}, {Lyytinen}, {Kajatkari},
  {Lehtinen}, {Markkanen}, \& {Toivari-Viitala}}]{Jet13}
{Jetsu}, L., {Porceddu}, S., {Lyytinen}, J., {et~al.} 2013, \apj, 773, 1

\bibitem[{{Kreiner}(2004)}]{Kre04}
{Kreiner}, J.~M. 2004, \actaa, 54, 207

\bibitem[{{Lehtinen} {et~al.}(2011){Lehtinen}, {Jetsu}, {Hackman}, {Kajatkari},
  \& {Henry}}]{Leh11}
{Lehtinen}, J., {Jetsu}, L., {Hackman}, T., {Kajatkari}, P., \& {Henry}, G.~W.
  2011, \aap, 527, A136

\bibitem[{{Li} {et~al.}(2018){Li}, {Rattenbury}, {Bond}, {Sumi}, {Bennett},
  {Koshimoto}, {Abe}, {Asakura}, {Barry}, {Bhattacharya}, {Donachie}, {Evans},
  {Fukui}, {Hirao}, {Itow}, {Masuda}, {Matsubara}, {Muraki}, {Nagakane},
  {Ohnishi}, {Saito}, {Sharan}, {Sullivan}, {Suzuki}, {Tristram}, \&
  {Yonehara}}]{Li18}
{Li}, M.~C.~A., {Rattenbury}, N.~J., {Bond}, I.~A., {et~al.} 2018, \mnras, 480,
  4557

\bibitem[{{Lomb}(1976)}]{Lom76}
{Lomb}, N.~R. 1976, \apss, 39, 447

\bibitem[{{Manzoori}(2016)}]{Man16}
{Manzoori}, D. 2016, Astronomy Letters, 42, 329

\bibitem[{{Mayo} {et~al.}(2019){Mayo}, {Rajpaul}, {Buchhave}, {Dressing},
  {Mortier}, {Zeng}, {Fortenbach}, {Aigrain}, {Bonomo}, {Collier Cameron},
  {Charbonneau}, {Coffinet}, {Cosentino}, {Damasso}, {Dumusque}, {Martinez
  Fiorenzano}, {Haywood}, {Latham}, {L{\'o}pez-Morales}, {Malavolta}, {Micela},
  {Molinari}, {Pearce}, {Pepe}, {Phillips}, {Piotto}, {Poretti}, {Rice},
  {Sozzetti}, \& {Udry}}]{May19}
{Mayo}, A.~W., {Rajpaul}, V.~M., {Buchhave}, L.~A., {et~al.} 2019, \aj, 158,
  165

\bibitem[{{Mellon} {et~al.}(2019){Mellon}, {Mamajek}, {Zwintz}, {David},
  {Stuik}, {Talens}, {Dorval}, {Burggraaff}, {Kenworthy}, {Bailey}, {Lomberg},
  {Kuhn}, {Ireland}, \& {Crawford}}]{Mel19}
{Mellon}, S.~N., {Mamajek}, E.~E., {Zwintz}, K., {et~al.} 2019, \apj, 870, 36

\bibitem[{{Murphy}(2012)}]{Mur12}
{Murphy}, S.~J. 2012, Astronomische Nachrichten, 333, 1057

\bibitem[{{Olspert} {et~al.}(2018){Olspert}, {Pelt}, {K{\"a}pyl{\"a}}, \&
  {Lehtinen}}]{Ols18}
{Olspert}, N., {Pelt}, J., {K{\"a}pyl{\"a}}, M.~J., \& {Lehtinen}, J. 2018,
  \aap, 615, A111

\bibitem[{{Pawlak} {et~al.}(2019){Pawlak}, {Pejcha}, {Jakub{\v{c}}{\'\i}k},
  {Jayasinghe}, {Kochanek}, {Stanek}, {Shappee}, {Holoien}, {Thompson},
  {Prieto}, {Dong}, {Shields}, {Pojmanski}, {Britt}, \& {Will}}]{Paw19}
{Pawlak}, M., {Pejcha}, O., {Jakub{\v{c}}{\'\i}k}, P., {et~al.} 2019, \mnras,
  487, 5932

\bibitem[{{Reinhold} \& {Reiners}(2013)}]{Rei13A}
{Reinhold}, T., \& {Reiners}, A. 2013, \aap, 557, A11

\bibitem[{{Scargle}(1982)}]{Sca82}
{Scargle}, J.~D. 1982, \apj, 263, 835

\bibitem[{{Zechmeister} \& {K{\"u}rster}(2009)}]{Zec09}
{Zechmeister}, M., \& {K{\"u}rster}, M. 2009, \aap, 496, 577
\end{thebibliography}

~


\appendix
\setcounter{table}{0}
\renewcommand{\thetable}{A\arabic{table}}
\renewcommand{\thefigure}{A\arabic{figure}}
\newcommand{\Ch}{\fbox{{\color{blue} 
\bf CHECK}}}
\setcounter{figure}{0}

\section{DCM computer code}

This appendix
gives the 
instructions for
the application of our
Discrete Chi-square Method
python
program \MyPro.
We also briefly
describe our Fisher test
program \PR{fisher.py}.
The user needs to copy
only the four files
\MyPro, \MyDat,
\MyTest ~and \PR{fisher.py}
from the
\Link{https://zenodo.org/}{Zenodo}
database.
These files should be stored
to the same directory.

We will update
these instructions
in our Zenodo database
\Link{https://zenodo.org/}
{Manual.}

\subsection{Control file}

The main idea is that the user
{\underline{\bf never edits}}
the \MyPro ~program,
but
{\underline{\bf only executes}}
it 
with the 
\PR{python \MyPro} command.
The user edits
{\underline{\bf only
the last right hand column}}
of the control file \MyDat.
This control file
\MyDat ~is shown
in the end of this appendix.
The \MyPro ~program may
stop working for these reasons:

\begin{enumerate}
  
\item Any \PR{``=''}
  character
  is removed from \MyDat,
  or added to  \MyDat.

\item Any of the first column 
  numbers \PR{1, 2, 3, ..., 24} is changed in \MyDat.
  The \MyPro ~program uses these
  numbers to identify the input values
  for the variables given
in the third column of \MyDat.

\item Values for variables
\PR{K1},
\PR{K2},
\PR{K3},
\PR{nL},
\PR{nS},
\PR{Rounds},
\PR{SimN}
and
\PR{SimRounds}
are not integers in \MyDat.

\end{enumerate}

Program \MyPro ~has
three different modes.
The \PR{SimMany} and \PR{RealData} values
in \MyDat ~determine these modes. 

\begin{center}
\begin{tabular}{lll}
\PR{SimMany $\neq$ 1}                             &
\PR{RealData = 1}                                 &  
Mode 1: \MyPro ~analyses one
sample of {\it real} data of \PR{file1}. \\ 
\PR{SimMany $\neq$ 1}                             &
\PR{RealData $\neq$ 1}                             &  
Mode 2: \MyPro ~creates and analyses one sample of {\it simulated} data.  \\ 
\PR{SimMany = 1}                                  &
Any \PR{RealData} value                           &  
                                                    Mode 3: \MyPro ~creates and analyses
 many samples of {\it simulated} data. \\
\end{tabular}
\end{center}

\noindent
Most users probably
  select Mode 1 for
  analysing the
  real data in 
  their own
  file \PR{file1}.
  This requires only the
  editing of
  lines 1-15
  in \MyDat.
Their
real data analysis
results
do not
depend on
the next lines 16-23
of \MyDat.
These variables
beginning 
with the letters \PR{Sim}
are relevant
only in the
simulation Modes 2 and 3.

\subsection{Control file
  variables}

In this section,
we use the same 
numbering 
of the control file \MyDat ~variables
as in this file itself.
The users can easily find the
description
of each variable, because 
their numbers below are 
highlighted with
yellow background.
This same numbering is also used in 
the figure and table
reproduction information
of Table
\ref{TableReproduce}.

\begin{enumerate}

\hlitem \PR{Tag} is text written
  to the beginning of the names
of all output figures and 
 files. 
This parameter \PR{Tag} allows 
the user to store the results
of each particular
\MyDat ~analysis
into figures and files having
the chosen
specific names. 
For our chosen
\PR{Tag = Dec2019} in \MyDat,
those output figures are

\subitem \PR{Dec2019z.eps}
(Fig. \ref{FigZ})

\subitem \PR{Dec2019gsim.eps}
(Only if
\PR{RealData $\neq$ 1}
and \PR{SimMany $\neq$ 1},
or
\PR{SimMany = 1}) 

\subitem \PR{Dec2019gdet.eps}
(Fig. \ref{FigModel})

\subitem \PR{Dec2019fA.eps}
(Fig. \ref{FigTwoSample})

\subitem \PR{Dec2019Many}
(Fig. \ref{FigMany}:
Only if \PR{SimMany = 1})

\item[] The output files are

  \subitem \PR{Dec2019Params.dat}
  (Analysis results:
  parts of Table \ref{TableSimOne})

  \subitem \PR{Dec2019Residuals.dat}
  (Residuals file)

  \subitem \PR{Dec2019Model.dat}
  (Model file)

  \subitem \PR{Dec2019AllBeta.dat}
  (Free parameter file)

  \subitem \PR{Dec2019ManyfA.dat }
  (Relative frequency error file:
  Only if \PR{SimMany = 1})
  
\item[] The contents of these output files
  are described later in this
  appendix.

\hlitem  \PR{RealData} is used
to select the analysed data.
Its value 
is relevant only in Modes 1 and 2
when \PR{SimMany $\neq$    1}.
  \subitem  \PR{RealData=1} activates the
  {\it \underline{Mode 1}}
  of program \MyPro,
where it analyses the
real data given in file \PR{file1}.

\subitem   \PR{RealData $\neq$ 1}
activates the
{\it \underline{Mode 2}}
of program \MyPro,
where it creates simulated data,
stores these data to
a file and analyses these data.
It also creates a figure of the
simulation model
and the simulated data.
For the \PR{Tag = Dec2019}
in our \MyDat,
the name of the input
data figure is
\PR{Dec2019gsim.eps}.
The simulated
and analysed data file is
\PR{Dec2019SimulatedData.dat}. 
The output figure is
\PR{Dec2019gdet.eps}.
  The user can test
  many problems
  encountered with real
  data by 
simulating data
having the same 
time span $\Delta T=$\PR{SimDT},
sample size $n^{\star}=$\PR{SimN} and
signal to noise ratio
${\mathrm{SN}}=$\PR{SimSN} as the real data.
For \PR{SimT $\neq$ 1}, the time
points for the simulated
data are the same 
as for the real data in \PR{file1}.
With our \PR{Tag=Dec2019},
the comparison between
simulated
\PR{Dec2019gsim.eps}
and
detected
\PR{Dec2019gdet.eps}
figures
reveals directly,
if DCM succeeds.
  
\hlitem \PR{file1} is the name
  of the file containing the 
  real data analysed in Mode 1
  (\PR{RealData = 1} and \PR{SimMany $\neq$ 1}).
  Its format must be the same
  as in our
  Table \ref{TableSimulatedData}.
  Our \PR{TestData.dat} file
  in Zenodo database contains
  the numerical values of
  Table \ref{TableSimulatedData}.
  In this paper, we analyse
  these 
  Table \ref{TableSimulatedData}
  simulated data
  with \MyPro,
  and show our results
  in
  Figs.
  \ref{FigZ}-\ref{FigTooFewThree}
  and Table \ref{TableModels}.
  All these results can be
  reproduced with the input
  variable values
  given in Table
  \ref{TableReproduce}.

\hlitem \PR{dummy} is the value for 
input and output which contains no
information. We use \PR{dummy=-99.999}.

\subitem None of the analysed
real or simulated
observations \PR{Y}$=y_i$ should have 
the numerical value of \PR{dummy}.

\subitem None of the model
parameters should have
the numerical value of  \PR{dummy}.

\hlitem \PR{K1}$=K_1=$  1, 2, 3, 4,
5 or 6 signals (Eq. \ref{Eqmodel})
  
\hlitem \PR{K2}$= K_2 = $ 1 or 2
  signal order (Eq. \ref{Eqmodel})
  
\hlitem \PR{K3}$= K_3 = $
  0, 1, 2, 3, 4,
  5 or 6 order
  polynomial trend
  (Eq. \ref{Eqmodel})
  
\hlitem \PR{nL}$= n_{\mathrm{L}}= $
  number of
 tested frequencies in long search

\hlitem \PR{nS}$= n_{\mathrm{L}}= $
  number 
  of tested frequencies
  in short search

\hlitem \PR{c}$=c=$ width
  of short tested 
  frequency interval
  (Eq. \ref{EqShort})

\hlitem \PR{TestStat} is used to
  select the test statistic
\PR{z}$=z$.

\subitem If  \PR{TestStat=1}, 
$z$ is computed
from $\chi^2$ (Eq. \ref{EqZone}:
data errors known). 

\subitem If  \PR{TestStat $\neq$ 1},
$z$ is computed
from $R$ (Eq. \ref{EqZtwo}:
data errors unknown).

\hlitem \PR{PMIN}$=P_{\mathrm{min}}=$
  minimum period
  
  \subitem If \PR{RealData=1}
and
  \PR{SimMany $\neq$ 1}
  (Mode 1), 
\PR{PMIN} is the minimum
tested period for the real
data in \PR{file1}.

\subitem
If \PR{RealData $\neq$ 1}
and
\PR{SimMany $\neq$ 1} (Mode 2),
or \PR{SimMany = 1} 
(Mode 3),
\PR{PMIN} is the minimum value
for the random periods
of simulated model(-s),
as well as the minimum
of tested periods for the
simulated data.

\hlitem \PR{PMAX}$=P_{\mathrm{max}}=$
  maximum period

  \subitem If \PR{RealData=1} and
  \PR{SimMany $\neq$ 1}
  (Mode 1), 
\PR{PMAX} is the maximum
tested period for the real data
in \PR{file1}.

\subitem If \PR{RealData $\neq$ 1}
and
  \PR{SimMany $\neq$ 1}
(Mode 2)
or \PR{SimMany = 1} 
(Mode 3),
\PR{PMAX} is the maximum value 
for the random periods
of simulated model(-s),
as well as the maximum
of tested periods for the
simulated data.

\hlitem \PR{Rounds}$=$ number of
  bootstrap rounds

\hlitem \PR{NonLinear} determines,
  if
  a non-linear model iteration
  is performed.

\subitem If \PR{NonLinear=1},
program \MyPro
~performs a nonlinear iteration 
from $\beta_{\mathrm{initial}}$
to $\beta_{\mathrm{final}}$
(Eq. \ref{EqInitial}). 

\subitem If \PR{NonLinear$\neq$1},
program
\MyPro
~does not perform a non-linear iteration,
and only uses
the $\beta_{\mathrm{initial}}$ value
of Eq. \ref{EqInitial}.
This alternative may 
cause error messages,
because low $n_{\mathrm{L}}$
and $n_{\mathrm{S}}$
test grid values can
give the same results
during many bootstrap rounds.

\hlitem \PR{SimT} determines the $t^{\star}_i$
time points of simulated data
when \PR{RealData $\neq 1$} and
\PR{SimMany $\neq$ 1} (Mode 2),
or \PR{SimMany = 1} (Mode 3).

  \subitem If \PR{SimT = 1}, these simulated
  $t_I^{\star}$
  time points are drawn from
  the uniform random distribution of
  Eq. \ref{EqRandomT}.
  \subitem If \PR{SimT $\neq$ 1}, the $t_i$
  time points of real data in file \PR{file1}
  are used as time points $t_i^{\star}$ in
  the simulations.
  In this alternative, the user
  can simulate data having
  same time points $t_i=$\PR{T} as the real data.
  The user can also adjust 
  the sample size \PR{SimN},
  the signal to noise ratio
  \PR{SimSN} and the time
  span \PR{SimDT} in \MyDat ~to
  the values of real data.

\hlitem \PR{SimN}$=n^{\star}=$ number of
  simulated observations when
  \PR{RealData $\neq$ 1} and
  \PR{SimMany $\neq$ 1} (Mode 2),
    or \PR{SimMany=1} (Mode 3). 
  
\hlitem  \PR{SimSN}$={\mathrm{SN}}=$
  signal to noise ratio
  of simulated observations
  (Eq. \ref{EqSigmaSim})
  when \PR{RealData $\neq$ 1}
  and \PR{SimMany $\neq$ 1} (Mode 2),
  or \PR{SimMany=1} (Mode 3).
  
\hlitem  \PR{SimDT}$=\Delta T$
  time span 
  of simulated observations
  when \PR{RealData $\neq$ 1}
  and \PR{SimMany $\neq$ 1} (Mode 2),
  or \PR{SimMany=1} (Mode 3).

\hlitem   \PR{SimMany} activates
  the {\it \underline{Mode 3}}
  of \MyPro.

\subitem If \PR{SimMany $\neq$ 1} 
and \PR{RealData=1}, 
\MyPro ~analyses {\it one} real
data sample (Mode 1).

\subitem If \PR{SimMany $\neq$ 1} 
and \PR{RealData $\neq$ 1},
\MyPro ~creates and analyses
{\it one} 
simulated data sample (Mode 2).

\subitem If \PR{SimMany = 1},
\MyPro ~creates and
analyses {\it many}, \PR{SimRounds},
simulated data samples
(Mode 3).
With our \PR{Tag = Dec2019} the
results are 
figure \PR{Dec2019Many.eps}
(Fig. \ref{FigMany})
and
file \PR{Dec2019AllBeta.dat}.

\hlitem  \PR{SimRounds} = Number of
  simulated data samples created
  and analysed when
  \PR{SimMany = 1}.

\hlitem  \PR{SimDF}$= f_{\mathrm{crit}}$ 
        of Eq. \ref{EqFreqCrit}
        (e.g. Fig. \ref{FigMany}:
        transparent
        diamonds)
        when \PR{SimMany = 1}.

\hlitem  \PR{SimDA}$= A_{\mathrm{crit}}$ 
        of Eq. \ref{EqAmpCrit}
        (e.g. Fig. \ref{FigMany}:
        transparent
        circles)
         when \PR{SimMany = 1}.

\hlitem  \PR{PrintScreen} controls
printing to screen in all Modes 1-3.

\subitem \PR{PrintScreen = 1} results are
printed to screen. This allows the user to
follow from the screen
how the computations proceed, like the
periods simulated and/or detected,
or the bootstrap rounds completed.

\subitem \PR{PrintScreen $\neq$ 1} prevents
  printing to screen. This may be
  needed, if \MyPro ~is executed in batch.

\end{enumerate}

\subsection{Analysis results file}

The main results of the analysis
are stored in the analysis results file.
The same information is also printed to
the screen, if \PR{PrintScreen = 1}. 
All values in our
Table \ref{TableSimOne} (Column 3) are
extracted from the result
file \PR{Dec2019Params.dat}
given in the end of this appendix.
The contents of this file
are summarized
in Table \ref{TableAnalysisResults}.

\subsection{Residuals file}

The $\epsilon_i$ residuals of the model
are stored into the residuals file,
which is
\PR{Dec2019Residuals.dat}
with our \PR{Tag = Dec2019}.
Columns 1-3 are
\PR{T}$=t_i$,
\PR{EPSILON}$=\epsilon_i=y_i-g_i$,
\PR{EY}$=\sigma_i$.
The format is the same as in the
real data file \PR{file1}, i.e.
the format of this 
residual file is
immediately
suitable
for further \MyPro ~period analysis.

\subsection{Model  file}

This file is
\PR{Dec2019Model.dat}
with our \PR{Tag = Dec2019}.
Columns 1-4 are
\PR{T}$=t_i$,
\PR{Y}$=y_i$,
\PR{EY}$=\sigma_i$
and
\PR{G}$=g_i$.

\subsection{Free parameter file}

This free parameter file is
\PR{Dec2019AllBeta.dat}
with our \PR{Tag = Dec2019}.
It contains 39 columns.
The first two columns are
\PR{TEPOCH}$=t_1$ and
\PR{DELTAT}$=\Delta T$.
For the free parameters $\beta_i$,
our \MyPro ~program uses the index values $i$
given in Table \ref{Tablebeta}. 
All columns having
\PR{dummy} values
contain no $\beta_i$ value.
If these \PR{dummy} columns are removed,
the remaining $i$ columns are
the $\beta_i$ free parameters specified
in Table \ref{Tablebeta}.
If necessary, the user can derive
all functions of the $g(t)$ model
from these
parameters $t_1, \Delta T$ and
$\beta_i$ by using Eqs. 
\ref{Eqmodel} - \ref{Eqpolynomial}.

\begin{itemize}

\item[] In Mode 1 (\PR{SimMany $\neq$ 1}, 
  \PR{RealData = 1}),
  the first line contains
  the parameters
  $t_1, \Delta T$ and
  $\beta_i$
  for the original data sample in \PR{file1}.
  Other lines contain these parameters
  for the bootstrap data samples.

\item[] In Mode 2
  (\PR{SimMany $\neq$ 1}, 
  \PR{RealData $\neq$ 1}),
  the first line contains the
  $t_1, \Delta T$ and
  $\beta_i$
  parameters
  for the original
  simulated data sample.
  Other lines contain
  these parameters
  for the bootstrap data samples.

\item[] In Mode 3
  (\PR{SimMany = 1},
  any \PR{RealData} value),
  the 1st, 3rd, ... ,
  $2 \times$\PR{SimRounds}$-1$
  lines contain the simulated
  $t_1, \Delta T$ and
   $\beta_i$ model parameters.
  Every next line,
  the 2nd, 4th, ... $2 \times$\PR{SimRounds} lines,
  contain the
  respective detected model
  parameters.

\end{itemize}

\subsection{Relative frequency error file}

This file is produced
only when \PR{SimMany = 1} (Mode 3).
With our \PR{Tag = Dec2019}, its name is
\PR{Dec2019ManyfA.dat}. 
Its contents are also printed in the
screen when \PR{PrintScreen = 1}.
This file gives the relative frequency
errors (Eq. \ref{EqRelativeF}).
The results are given separately for all models,
as well as for those models
that {\it do not} fulfill criteria
of Eqs. \ref{EqFreqCrit}
and \ref{EqAmpCrit},
like models not highlighted in
Fig. \ref{FigMany}.

We show
\PR{\MyDat} and
\PR{Dec2019Params.dat} files
in the end of this appendix, but not
\PR{Dec2019Residuals.dat},
\PR{Dec2019Model.dat},
\PR{Dec2019AllBeta.dat} and
\PR{Dec2019ManyfA.dat} files,
because they
can be created with the
\PR{python dcm.py} command.

\subsection{Reproducing
  our results}

Here, we explain how
the users can
reproduce our main results,
and at
the same time practice the
use of our CDM program \MyPro.
After
reproducing our
results,
the users can
also be
more confident
about results for
their own data.

\begin{itemize}

\item[] Col. 1 of Table \ref{TableReproduce} 
gives the input parameter values in
\PR{\MyDat} that reproduce
the results in our
Figs. \ref{FigZ}-\ref{FigTwoSample}.
File \MyDat ~from Zenodo database
contains
the same input values as Col. 1
of Table \ref{TableReproduce}. 

\item[] Col. 2 of
Table \ref{TableReproduce}  shows
how the results in our Table \ref{TableModels}
can be obtained by adjusting
the
\PR{K1}$=K_1$, 
\PR{K2}$=K_2$ and 
\PR{K3}$=K_3$
values in \MyDat. 
Those particular adjusted values
are denoted with * in Table \ref{TableReproduce}.

\item[]
Cols. 3-4 contain the \PR{\MyDat} input values
for reproducing
Figs. \ref{FigTooManyOne} -
\ref{FigTooFewThree}.

\item[]
Col. 5 tells how to reproduce
Fig. \ref{FigMany}.

\item[] Col. 6 shows how 
  Table \ref{TableFreqError}
  can be reproduced by adjusting
  \PR{SimN} and \PR{SimSN} values
  marked with ``*''.

\end{itemize}

\noindent
The users' results for
Figs. \ref{FigZ},
\ref{FigModel},
\ref{FigTooManyOne},
\ref{FigTooManyTwo},
\ref{FigTooFewOne} and
\ref{FigTooFewTwo}
should be identical.
However, the results for Figs.
\ref{FigTwoSample},
\ref{FigTooManyThree} and
\ref{FigTooFewThree},
as well as the error estimates
in Table \ref{TableSimOne},
will never be identical,
because the random bootstrap 
residuals $\epsilon_i^{\star}$
are always different
in Eq. \ref{EqBoot}.
The results in Fig. \ref{FigMany}
and Table \ref{TableFreqError}
will also always differ,
because the created random data
samples $y_i^{\star}$ of
Eq. \ref{EqSimManyData}
are never the same.

\subsection{Fisher test}

\begin{figure}
\begin{center}
  \resizebox{6.5cm}{!}
  {\includegraphics{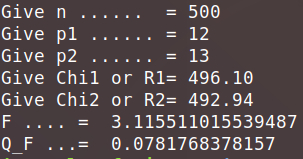}}
\end{center}
\caption{Application example of
  \PR{fisher.py}.}
\label{FigFisher}
\end{figure}

In this section,
we explain how our
program \PR{fisher.py}
computes the $F_{\chi}$ and $Q_F$
estimates in 
Table \ref{TableModels}.
Executing
\PR{python fisher.py}
asks for
the numerical input
values for \PR{n}$=n$, \PR{p1}$=p_1$,
\PR{p2}$=p_2$, \PR{Chi1}$=\chi_1$
or \PR{R1}$=R_1$,
and \PR{Chi2}$=\chi_2$ or \PR{R1}$=R_2$.
If the pair $\chi_1$ and $\chi_2$ is used,
the \PR{F}$=F_{\chi}$ value is computed
from Eq. \ref{EqFChi}.
For the $R_1$ and $R_2$ pairs,
the \PR{F}$=F_{R}$ value is computed
from Eq. \ref{EqFR}.
The program prints the \PR{F} value and
the \PR{Q}$=Q_F$ value of Eq. \ref{EqQF}.
The results for comparing models 19 and 20
are shown in Fig. \ref{FigFisher}.

\subsection{Least squares fit subroutine}

The numerical python least
squares subroutine 
\PR{optimize.leastsq}
minimizes the sum of squares
$\sum_{i=1}^n [(y_i-g_i)/\sigma_i]^2$. 
For the 32 nested models
of Table \ref{TableModels},
the $\chi^2$ estimates
for the linear models
of denser tested
frequency grids agree
with the results
for
the non-linear
model (Eq. \ref{EqInitial}).
In the python
\PR{optimize.leastsq} 
subroutine,
the
parameter \PR{ftol} 
measures the
relative error in the above sum of squares.
The parameter \PR{xtol}
measures the relative error in the
desired approximate solution.
We use 
\PR{ftol=0.0001} and
\PR{xtol=0.0001} in
\PR{optimize.leastsq}
of our non-linear
least squares fit
\PR{NonLinearLSF},
because this
should prevent
the numerical
non-linear iteration
$\bar{\beta}_{\mathrm{final}}$
of Eq. \ref{EqInitial}
from wandering too far
from the
unambiguous initial
$\bar{\beta}_{\mathrm{initial}}$
estimate obtained
from linear modelling.

Amplitude dispersion 
occurs already for the
unambiguous linear
models, {\it before} any
non-linear
modelling is made.
The numerical
\PR{optimize.leastsq} subroutine
has to utilize
these unrealistic
high amplitude $h_i(t)$
curves, because this is the only
possible way to minimize
$\chi^2$.
These high amplitude
curves, which nearly
cancel out each other, offer
the only possible way to
fit two or more curves
having nearly
the same frequencies.
From the purely mathematical
point of view,
these amplitude dispersion
models do not fail.
They are just unrealistic.
There just are no reasonable
low amplitude solutions.
However, we know for certain
that
these unrealistic 
models fail,
because we know
that model 19 is the correct solution.

The \PR{optimize.leastsq}
is not an analytical subroutine,
because it
does not require 
the model partial
derivative formulas
as its input.
Here is room
for development
for those who are 
prepared to code
these
$\partial g(t)/ \partial \beta_i$
partial
derivatives.
But even that 
analytical solution
could not eliminate
amplitude dispersion,
because the continuous and stable
$z_1(f_1)$, ..., $z_{K_1}(f_{K_1})$
periodograms already confirm
that there simply are no
realistic low amplitude
$h_i(t)$ curve solutions
(e.g. Figs. \ref{FigTooManyOne}
and \ref{FigTooFewOne}).

\subsection{Qualitative
  program code description}

We end this appendix with a short qualitative
description of
the stages of \MyPro.

\begin{enumerate}

\item First, the {\it long} tested frequency
  interval
  $f_{\mathrm{min}}=P_{\mathrm{max}}^{-1}$
  and $f_{\mathrm{max}}=P_{\mathrm{min}}^{-1}$ is fixed.
We create the $n_{\mathrm{L}}$ evenly spaced
  tested frequencies
  $f_1, f_2, ..., f_{K_1}$
  between $f_{\mathrm{min}}$ and  $f_{\mathrm{max}}$.
  Such grids are illustrated 
  in Fig. \ref{FigGrid}.
  The $j$:th tested value of frequency $f_i$
  is denoted with $f_{i,j_i}$
  $(j_i=1, 2, ..., n_{\mathrm{L}})$.
\label{Stageone}

\item 
We create the one-dimensional vectors 
$F_1,F_2,... F_{K_1},Z$
for collecting the period search
results from the $K_1$-dimensional
tested frequency space. 
These vectors are empty before the
loop of tested frequencies begins.
\label{Stagetwo}

\item All $f_{1,j_1} > f_{2,j_2} > ... >
  f_{K_1,j_{K_1}}$ combinations are tested
in a loop. For each combination,
\label{Stagethree}

\subitem (a) We compute the periodogram
test statistic
$z=z[f_{1,j_1}, f_{2,j_2}, ..., f_{K_1,j_{K_1}}]$
of Eq. \ref{EqZone} or \ref{EqZtwo}.

\subitem (b) The tested
frequency combination and
the result for $z$ are appended 
into the collection vectors
\begin{eqnarray}
F_{\mathrm{1,old}}  \leftarrow f_{1,j_1}   &
\Rightarrow                        & 
F_{\mathrm{1,new}}=[F_{\mathrm{1,old}},f_{1,j_1}]  
\nonumber                          \\
F_{\mathrm{2,old}}  \leftarrow f_{2,j_2}   & 
\Rightarrow                        & 
F_{\mathrm{2,new}}=[F_{\mathrm{2,old}},f_{2,j_2}] 
\nonumber                          \\
  ...   
                                   & 
\Rightarrow                        & 
  ...         \label{Eqproject}    \\
F_{\mathrm{K_1,old}}\leftarrow f_{K_1,j_{K_1}}  &
\Rightarrow                        & 
F_{\mathrm{K_1,new}}=[F_{\mathrm{K_1,old}},f_{K_1,j_{K_1}}]   
\nonumber                          \\
Z_{\mathrm{old}}   \leftarrow z        &
\Rightarrow                        & 
  Z_{\mathrm{new}}=[Z_{\mathrm{old}},z]                      \nonumber 
\end{eqnarray}

\item After completing
  the test frequency loop,
 we find the index $k$ of the $Z$ minimum.
  The best frequency candidates are at
  $F_1[k]=f_{\mathrm{1,mid}}$,
  $F_2[k]=f_{\mathrm{2,mid}}$, ...
  $F_{K_1}[k]=f_{\mathrm{K_1,mid}}$
  of Eq. \ref{EqShort}
(e.g. Fig.~\ref{FigGrid}: diamonds).
\label{Stagefour}

\item We fix the {\it short}
  denser tested frequency
  grids of Eq. \ref{EqShort},
  which are
centered at 
$F_1[k]$,
$F_2[k]$, ... $F_{K_1}[k]$.
The more accurate best frequency 
values are determined by
using these dense grids 
of $n_{\mathrm{S}}$ tested
frequencies.

\item The bootstrap is used to solve 
the errors for model parameters
within the {\it short} tested
frequency
intervals.

\end{enumerate}

\noindent
The tested 
frequency combination
$\bar{\beta}_{I}=
[f_{1,j_1}, f_{2,j_2}, ..., f_{K_1,j_  {K_1}}]$ 
is swapped before stage \ref{Stagethree}a.
We write the next tested frequencies into file
\PR{ALLF.dat} with subroutine \PR{WriteALLF}.
Subroutine \PR{ReadALLF} 
reads these frequencies
within another subroutine \PR{LinearModel}.
Hence, we do not have to rewrite
the model equations for every new tested 
$\bar{\beta}_{I}=
[f_{1,j_1}, f_{2,j_2}, ..., f_{K_1,j_{K_1}}]$ combination.
In stage \ref{Stagethree}b,
the results from the $K_1$-dimensional 
frequency grid space are projected into the
one-dimensional collection
vectors of Eq. \ref{Eqproject}.

We emphasize that although there 
may be coding errors in our
\PR{dcm.py} program,
all our main conclusions apply.
Ours is just
one possible DCM
application code.
More talented coders
can certainly improve 
our code.

%
%
%
\begin{table}
  \caption{Control file
    variables reproducing our results.
    Columns 1-6 are explained
    in
    Section
    ``Reproducing our results''.}
\begin{center}
\begin{tabular}{rlcccccc}
  \hline
 & Reproduce:    &
                 Figs.  \ref{FigZ}, \ref{FigModel}
                 and \ref{FigTwoSample}   
               &  Table \ref{TableModels}
               & Figs.  \ref{FigTooManyOne}, \ref{FigTooManyTwo}
                 and \ref{FigTooManyThree}
               & Figs.  \ref{FigTooFewOne}, \ref{FigTooFewTwo}
                 and \ref{FigTooFewThree}
               & Fig. \ref{FigMany}
               &  Table \ref{TableFreqError}
  \\
\hline
   &                  & Col. 1         & Col. 2           & Col. 3      & Col. 4        & Col. 5       & Col. 6 \\
  \hline
 1 & \PR{Tag}         & Dec2019        & Dec2019          & Dec2019     & Dec2019       & Dec2019      & Dec2019      \\
 2 & \PR{RealData}    & 1              & 1                & 1           & 1             & 1            & 1            \\
 3 & \PR{file1}       & TestData.dat   & TestData.dat     & TestData.dat& TestData.dat  & TestData.dat & TestData.dat \\
 4 & \PR{dummy}       & -99.999        & -99.999          & -99.999     & -99.999       & -99.999      & -99.999      \\
 5 & \PR{K1}          & 3              & *                & 4           & 2             & 3            & 3            \\
 6 & \PR{K2}          & 1              & *                & 1           & 1             & 1            & 1            \\
 7 & \PR{K3}          & 2              & *                & 2           & 0             & 1            & 1            \\
 8 & \PR{nL}          & 60             & 60               & 60          & 60            & 60           & 60           \\
 9 & \PR{nS}          & 30             & 30               & 30          & 30            & 30           & 30           \\
10 & \PR{c}           & 0.20           & 0.20             & 0.20        & 0.20          & 0.20         & 0.20         \\
11 & \PR{TestStat}    & 1              & 1                & 1           & 1             & 1            & 1            \\
12 & \PR{PMIN}        & 1.0            & 1.0              & 1.0         & 1.0           & 1.0          & 1.0          \\
13 & \PR{PMAX}        & 2.0            & 2.0              & 2.0         & 2.0           & 2.0          & 2.0          \\
14 & \PR{Rounds}      & 30             & 2                & 30          & 50            & 3            & 3            \\
15 & \PR{NonLinear}   & 1              & 1                & 1           & 1             & 1            & 1            \\
16 & \PR{SimT}        & 1              & 1                & 1           & 1             & 1            & 1            \\
17 & \PR{SimN}        & 500            & 500              & 500         & 500           & 500          & *            \\
18 & \PR{SimSN}       & 100            & 100              & 100         & 100           & 100          & *            \\
19 & \PR{SimDT}       & 4.0            & 4.0              & 4.0         & 4.0           & 4.0          & 4.0          \\
20 & \PR{SimMany}     & 0              & 0                & 0           & 0             & 1            & 1            \\
21 & \PR{SimRounds}   & 3              & 3                & 3           & 3             & 30           & 100          \\
22 & \PR{SimDF}       & 0.05           & 0.05             & 0.05        & 0.05          & 0.05         & 0.05         \\
23 & \PR{SimDA}       & 0.5            & 0.5              & 0.5         & 0.5           & 0.5          & 0.5          \\
24 & \PR{PrintScreen} & 1.0            & 1.0              & 1.0         & 1.0           & 1.0          & 1.0          \\
\hline
\end{tabular}
\end{center}
\label{TableReproduce}
\end{table}

\begin{table}
  \caption{Parameters of result file.}
  \begin{center}
\begin{tabular}{ccl}
\hline
  \PR{n,T1,DT}  & $n,t_1,\Delta T$             & number of observations, first observing time, time span \\
\PR{my,sy,SN}   & $m_y,s_y,{\mathrm{SN}}$       & mean, standard deviation and signal to noise ratio of observations\\
\PR{K1,K2,K3}   & $K_1,K_2,K_3$                 & number of signals, signal order and polynomial trend order \\
\PR{p}          & $p$                          & number of free parameters    \\
\PR{PMIN,PMAX}  & $P_{\mathrm{min}},P_{\mathrm{max}}$  & minimum and maximum tested period \\
\PR{nL,nS}      & $n_{\mathrm{L}},n_{\mathrm{S}}$     & number of tested frequencies in long and short search \\
\PR{CHI2,R}     & $\chi^2,R$                   & chi-square and sum of squared residuals of the best model \\
\PR{F1,P1,A1,T1MIN1,T1MIN2,T1MAX1,T1MAX2}      & $h_1(t)$ & parameters $f_1, P_1, A_1,
t_{\mathrm{1,min,1}}
t_{\mathrm{1,min,2}}
t_{\mathrm{1,max,1}},
t_{\mathrm{1,max,2}}$ \\
\PR{F2,P2,A2,T2MIN1,T2MIN2,T2MAX1,T2MAX2}      & $h_2(t)$ & parameters $f_2, P_2, A_2,
t_{\mathrm{2,min,1}}
t_{\mathrm{2,min,2}}
t_{\mathrm{2,max,1}},
t_{\mathrm{2,max,2}}$ \\
\PR{F3,P3,A3,T3MIN1,T3MIN2,T3MAX1,T3MAX2}     & $h_3(t)$ & parameters $f_3, P_3, A_3,
t_{\mathrm{3,min,1}}
t_{\mathrm{3,min,2}}
t_{\mathrm{3,max,1}},
t_{\mathrm{3,max,2}}$ \\
\PR{BETA[i]} & $\bar{\beta}$ & all free parameter values \\
\hline
\end{tabular}
\end{center}
\label{TableAnalysisResults}
\end{table}

\begin{table*}
  \caption{Programming indexes $i$ for
    free parameters $\beta_i$. 
These indexes are given separately 
for $\bar{\beta}_{I}$ and
$\bar{\beta}_{II}$. Columns ``L'' and ``N''
denote Linear and Non-linear models
for given $K_1$ and $K_2$ combinations.  }
\begin{small}
\addtolength{\tabcolsep}{-0.08cm}
\begin{center}
\begin{tabular}{ccccccccccccccccccccccccccccccc}
\hline
 & \multicolumn{29}{c}{$\bar{\beta}_{I}$ free parameters} & \\
\cline{2-30}
         & \multicolumn{4}{c}{One period} & 
         & \multicolumn{4}{c}{Two periods} & 
         & \multicolumn{4}{c}{Three periods} & 
         & \multicolumn{4}{c}{Four periods} & 
         & \multicolumn{4}{c}{Five periods} & 
         & \multicolumn{4}{c}{Six periods} \\
\cline{2-5}
\cline{7-10}
\cline{12-15}
\cline{17-20}
\cline{22-25}
\cline{27-30}
         & \multicolumn{2}{c}{$K_1=1$} & \multicolumn{2}{c}{$K_1=1$} &
         & \multicolumn{2}{c}{$K_1=2$} & \multicolumn{2}{c}{$K_1=2$} &
         & \multicolumn{2}{c}{$K_1=3$} & \multicolumn{2}{c}{$K_1=3$} &
         & \multicolumn{2}{c}{$K_1=4$} & \multicolumn{2}{c}{$K_1=4$} &
         & \multicolumn{2}{c}{$K_1=5$} & \multicolumn{2}{c}{$K_1=5$} &
         & \multicolumn{2}{c}{$K_1=6$} & \multicolumn{2}{c}{$K_1=6$} & \\
         & \multicolumn{2}{c}{$K_2=1$} & \multicolumn{2}{c}{$K_2=2$} &
         & \multicolumn{2}{c}{$K_2=1$} & \multicolumn{2}{c}{$K_2=2$} &
         & \multicolumn{2}{c}{$K_2=1$} & \multicolumn{2}{c}{$K_2=2$} &
         & \multicolumn{2}{c}{$K_2=1$} & \multicolumn{2}{c}{$K_2=2$} &
         & \multicolumn{2}{c}{$K_2=1$} & \multicolumn{2}{c}{$K_2=2$} &
         & \multicolumn{2}{c}{$K_2=1$} & \multicolumn{2}{c}{$K_2=2$} & \\
\hline
         &L&N&L&N&&L&N&L&N&&L&N&L&N&&L&N&L&N&&L&N&L&N&&L&N&L&N& \\
\hline
$f_1$    &-&1&-&1&&-&1&-&1&&-&1&-&1&&-&1&-&1&&-&1&-&1&&-&1&-&1& \\
$f_2$    &-&-&-&-&&-&2&-&2&&-&2&-&2&&-&2&-&2&&-&2&-&2&&-&2&-&2& \\ 
$f_3$    &-&-&-&-&&-&-&-&-&&-&3&-&3&&-&3&-&3&&-&3&-&3&&-&3&-&3& \\
$f_4$    &-&-&-&-&&-&-&-&-&&-&-&-&-&&-&4&-&4&&-&4&-&4&&-&4&-&4& \\
$f_5$    &-&-&-&-&&-&-&-&-&&-&-&-&-&&-&-&-& &&-&5&-&5&&-&5&-&5& \\
$f_6$    &-&-&-&-&&-&-&-&-&&-&-&-&-&&-&-&-&-&&-& -&-&-&&-&6&-&6& \\
\hline
 & \multicolumn{29}{c}{$\bar{\beta}_{II}$ free parameters} & \\
\cline{2-30}
         & \multicolumn{4}{c}{One period} & 
         & \multicolumn{4}{c}{Two periods} & 
         & \multicolumn{4}{c}{Three periods} & 
         & \multicolumn{4}{c}{Four periods} & 
         & \multicolumn{4}{c}{Five periods} & 
         & \multicolumn{4}{c}{Six periods} \\
\cline{2-5}
\cline{7-10}
\cline{12-15}
\cline{17-20}
\cline{22-25}
\cline{27-30}
         & \multicolumn{2}{c}{$K_1=1$} & \multicolumn{2}{c}{$K_1=1$} &
         & \multicolumn{2}{c}{$K_1=2$} & \multicolumn{2}{c}{$K_1=2$} &
         & \multicolumn{2}{c}{$K_1=3$} & \multicolumn{2}{c}{$K_1=3$} &
         & \multicolumn{2}{c}{$K_1=4$} & \multicolumn{2}{c}{$K_1=4$} &
         & \multicolumn{2}{c}{$K_1=5$} & \multicolumn{2}{c}{$K_1=5$} &
         & \multicolumn{2}{c}{$K_1=6$} & \multicolumn{2}{c}{$K_1=6$} & \\
         & \multicolumn{2}{c}{$K_2=1$} & \multicolumn{2}{c}{$K_2=2$} &
         & \multicolumn{2}{c}{$K_2=1$} & \multicolumn{2}{c}{$K_2=2$} &
         & \multicolumn{2}{c}{$K_2=1$} & \multicolumn{2}{c}{$K_2=2$} &
         & \multicolumn{2}{c}{$K_2=1$} & \multicolumn{2}{c}{$K_2=2$} &
         & \multicolumn{2}{c}{$K_2=1$} & \multicolumn{2}{c}{$K_2=2$} &
         & \multicolumn{2}{c}{$K_2=1$} & \multicolumn{2}{c}{$K_2=2$} & \\
\hline
         &L&N&L&N&&L&N&L&N&&L&N&L&N&&L&N&L&N&&L&N&L&N&&L&N&L&N& \\
\hline
$B_{1,1}$ &1&2&1&2&& 1&3& 1& 3&&1&4& 1& 4&&1& 5& 1& 5&& 1& 6& 1& 6&& 1& 7& 1& 7& \\
$C_{1,1}$ &2&3&2&3&& 2&4& 2& 4&&2&5& 2& 5&&2& 6& 2& 6&& 2& 7& 2& 7&& 2& 8& 2& 8& \\
$B_{1,2}$ &-&-&3&4&& -&-& 3& 5&&-&-& 3& 6&&-& -& 3& 7&& -& -& 3& 8&& -& -& 3& 9& \\
$C_{1,2}$ &-&-&4&5&& -&-& 4& 6&&-&-& 4& 7&&-& -& 4& 8&& -& -& 4& 9&& -& -& 4&10& \\
\hline
$B_{2,1}$ &-&-&-&-&& 3&5& 5& 7&&3&6& 5& 8&&3& 7& 5& 9&& 3& 8& 5&10&& 3& 9& 5&11& \\
$C_{2,1}$ &-&-&-&-&& 4&6& 6& 8&&4&7& 6& 9&&4& 8& 6&10&& 4& 9& 6&11&& 4&10& 6&12& \\
$B_{2,2}$ &-&-&-&-&&- &-& 7& 9&&-&-& 7&10&&-& -& 7&11&& -& -& 7&12&& -& -& 7&13& \\
$C_{2,2}$ &-&-&-&-&&- &-& 8&10&&-&-& 8&11&&-& -& 8&12&& -& -& 8&13&& -& -& 8&14& \\
\hline
$B_{3,1}$ &-&-&-&-&& -&-& -& -&&5&8& 9&12&&5& 9& 9&13&& 5&10& 9&14&& 5&11& 9&15& \\
$C_{3,1}$ &-&-&-&-&& -&-& -& -&&6&9&10&13&&6&10&10&14&& 6&11&10&15&& 6&12&10&16& \\
$B_{3,2}$ &-&-&-&-&&- &-& -& -&&-&-&11&14&&-& -&11&15&& -& -&11&16&& -& -&11&17& \\
$C_{3,2}$ &-&-&-&-&&- &-& -&- &&-&-&12&15&&-& -&12&16&& -& -&12&17&& -& -&12&18& \\
\hline
$B_{4,1}$ &-&-&-&-&& -&-& -& -&&-&-& -&- &&7&11&13&17&& 7&12&13&18&& 7&13&13&19& \\
$C_{4,1}$ &-&-&-&-&& -&-& -& -&&-&-& -&- &&8&12&14&18&& 8&13&14&19&& 8&14&14&20& \\
$B_{4,2}$ &-&-&-&-&&- &-& -& -&&-&-& -&- &&-&- &15&19&& -& -&15&20&& -& -&15&21& \\
$C_{4,2}$ &-&-&-&-&&- &-& -&- &&-&-& -&- &&-&- &16&20&& -& -&16&21&& -& -&16&22& \\
\hline
$B_{5,1}$ &-&-&-&-&& -&-& -& -&&-&-& -&- &&-& -& -& -&& 9&14&17&22&& 9&15&17&23& \\
$C_{5,1}$ &-&-&-&-&& -&-& -& -&&-&-& -&- &&-& -& -& -&&10&15&18&23&&10&16&18&24& \\
$B_{5,2}$ &-&-&-&-&&- &-& -& -&&-&-& -&- &&-&- & -& -&& -& -&19&24&& -& -&19&25& \\
$C_{5,2}$ &-&-&-&-&&- &-& -&- &&-&-& -&- &&-&- & -& -&& -& -&20&25&& -& -&20&26& \\
\hline
$B_{6,1}$ &-&-&-&-&& -&-& -& -&&-&-& -&- &&-& -& -& -&& -& -& -& -&&11&17&21&27& \\
$C_{6,1}$ &-&-&-&-&& -&-& -& -&&-&-& -&- &&-& -& -& -&& -& -& -& -&&12&18&22&28& \\
$B_{6,2}$ &-&-&-&-&&- &-& -& -&&-&-& -&- &&-&- & -& -&& -& -& -& -&& -& -&23&29& \\
$C_{6,2}$ &-&-&-&-&&- &-& -&- &&-&-& -&- &&-&- & -& -&& -& -& -& -&& -& -&24&30& \\
\hline
$M_0$    &3& 4& 5& 6&& 5& 7& 9&11&& 7&10&13&16&& 9&13&17&21&&11&16&21&26&&13&19&25&31& \\
$M_1$    &4& 5& 6& 7&& 6& 8&10&12&& 8&11&14&17&&10&14&18&22&&12&17&22&27&&14&20&26&32& \\
$M_2$    &5& 6& 7& 8&& 7& 9&11&13&& 9&12&15&18&&11&15&19&23&&13&18&23&28&&15&21&27&33& \\
$M_3$    &6& 7& 8& 9&& 8&10&12&14&&10&13&16&19&&12&16&20&24&&14&19&24&29&&16&22&28&34& \\
$M_4$    &7& 8& 9&10&& 9&11&13&15&&11&14&17&20&&13&17&21&25&&15&20&25&30&&17&23&29&35& \\
$M_5$    &8& 9&10&11&&10&12&14&16&&12&15&18&21&&14&18&22&26&&16&21&26&31&&18&24&30&36& \\
$M_6$    &9&10&11&12&&11&13&15&17&&13&16&19&22&&15&19&23&27&&17&22&27&32&&19&25&31&37& \\
\hline
\end{tabular}
\end{center}
\end{small}
\addtolength{\tabcolsep}{+0.08cm}
\label{Tablebeta}
\end{table*}

\clearpage

%
%
\begin{center}
Contents of \MyDat ~ \vspace{-0.5cm} 
\end{center}

\begin{scriptsize}
  {\color{magenta}
\begin{verbatim}
 1  = Tag         = Dec2019
 2  = RealData    = 1
 3  = file1       = TestData.dat
 4  = dummy       = -99.999
 5  = K1          = 3
 6  = K2          = 1
 7  = K3          = 2
 8  = nL          = 60
 9  = nS          = 30
10  = c           = 0.20
11  = TestStat    = 1
12  = PMIN        = 1.0
13  = PMAX        = 2.0
14  = Rounds      = 30
15  = NonLinear   = 1
16  = SimT        = 1
17  = SimN        = 500
18  = SimSN       = 100
19  = SimDT       = 4.0
20  = SimMany     = 0
21  = SimRounds   = 3
22  = SimDF       = 0.05
23  = SimDA       = 0.50
24  = PrintScreen = 1.0
\end{verbatim}
}
\end{scriptsize}

\begin{center}
Contents of \PR{Dec2019Params.dat} ~ \vspace{-0.3cm} 
\end{center}
\begin{scriptsize}
{\color{magenta}
\begin{verbatim}
                                          n              500
                                         T1 1.9547820000e-03
                                         DT 3.9958894740e+00
                                        my -1.2567031722e+00
                                         sy 2.2066570834e+00
                                      sigma 2.6888315618e-02
                                         SN 4.6424393691e+02
                                         K1                3
                                         K2                1
                                         K3                2
                                          p               12
                                       PMIN 1.0000000000e+00
                                       PMAX 2.0000000000e+00
                                         nL               60
                                         nS               40
         CHI2 4.9309212497e+02   gives ZMIN 9.9306809934e-01
            R 5.6396874104e-01   gives ZMIN 3.3584780513e-02
.................................................................
                    F1 9.0911731600e-01 +/- 1.2735271893e-03
                    P1 1.0999680486e+00 +/- 1.5420877320e-03
                    A1 9.0070629802e-01 +/- 1.5914601221e-02     SN      67.00
                T1MIN1 3.2534538828e-01 +/- 1.1847765288e-03
                T1MIN2              ... +/-             ...
                T1MAX1 8.7532941256e-01 +/- 5.3225925012e-04
                T1MAX2              ... +/-             ...
.................................................................
                    F2 7.1406875500e-01 +/- 6.6356730224e-03
                    P2 1.4004253694e+00 +/- 1.3149673657e-02
                    A2 1.0016254884e+00 +/- 3.2552250810e-02     SN      74.50
                T2MIN1 4.9569244560e-02 +/- 1.4268052938e-02
                T2MIN2              ... +/-             ...
                T2MAX1 7.4978192926e-01 +/- 7.7035844527e-03
                T2MAX2              ... +/-             ...
.................................................................
                    F3 5.2638452700e-01 +/- 4.0181847831e-03
                    P3 1.8997518899e+00 +/- 1.4379538506e-02
                    A3 1.1011510476e+00 +/- 4.6003971393e-02     SN      81.91
                T3MIN1 4.2559945346e-01 +/- 9.8085696951e-03
                T3MIN2              ... +/-             ...
                T3MAX1 1.3754753984e+00 +/- 2.6765025920e-03
                T3MAX2              ... +/-             ...
...............................................................
                     i              BETA[i]
                     1         F1  9.09117e-01 1.27353e-03 
                     2         F2  7.14069e-01 6.63567e-03 
                     3         F3  5.26385e-01 4.01818e-03 
                     4        B11  1.22518e-01 2.30736e-03 
                     5        C11 -4.33368e-01 8.68889e-03 
                     6        B21 -4.89585e-01 2.14509e-02 
                     7        C21 -1.05456e-01 3.02140e-02 
                     8        B31 -9.37527e-02 2.12900e-02 
                     9        C31 -5.42535e-01 2.62038e-02 
                    10         M0  1.79955e+00 2.20915e-03 
                    11         M1 -1.49889e+00 2.81791e-03 
                    12         M2 -1.20073e+00 1.33342e-03 
\end{verbatim}
}
\end{scriptsize}

\end{document}